\input amstex
\input amssym.tex
\input amssym
%
%
%
%
\def\temp{1.31}
\let\tempp=\relax
\expandafter\ifx\csname psboxversion\endcsname\relax
  \message{version: \temp}
\else
    \ifdim\temp cm>\psboxversion cm
      \message{version: \temp}
    \else
      \message{psbox(\psboxversion) is already loaded: I won't load
        psbox(\temp)!}
      \let\temp=\psboxversion
      \let\tempp= 
    \fi
\fi
\tempp
\let\psboxversion=\temp
\catcode`\@=11
%
%
\def\execute#1{#1}
\def\psm@keother#1{\catcode`#112\relax}
\def\executeinspecs#1{%
\execute{\begingroup\let\do\psm@keother\dospecials\catcode`\^^M=9#1\endgroup}}
%
%
\def\psfortextures{
\def\PSspeci@l##1##2{%
\special{illustration ##1\space scaled ##2}%
}}
\def\psfordvitops{
\def\PSspeci@l##1##2{%
\special{dvitops: import ##1\space \the\drawingwd \the\drawinght}%
}}
\def\psfordvips{
\def\PSspeci@l##1##2{%
\d@my=0.1bp \d@mx=\drawingwd \divide\d@mx by\d@my%
\includegraphics{##1\space}%
}}
\def\psforoztex{
\def\PSspeci@l##1##2{%
\special{##1 \space
      ##2 1000 div dup scale
      \putsp@ce{\number-\psllx} \putsp@ce{\number-\pslly} translate
}%
}}
\def\putsp@ce#1{#1 }
\def\psfordvitps{
\def\psdimt@n@sp##1{\d@mx=##1\relax\edef\psn@sp{\number\d@mx}}
\def\PSspeci@l##1##2{%
\special{dvitps: Include0 "psfig.psr"}
\psdimt@n@sp{\drawingwd}
\special{dvitps: Literal "\psn@sp\space"}
\psdimt@n@sp{\drawinght}
\special{dvitps: Literal "\psn@sp\space"}
\psdimt@n@sp{\psllx bp}
\special{dvitps: Literal "\psn@sp\space"}
\psdimt@n@sp{\pslly bp}
\special{dvitps: Literal "\psn@sp\space"}
\psdimt@n@sp{\psurx bp}
\special{dvitps: Literal "\psn@sp\space"}
\psdimt@n@sp{\psury bp}
\special{dvitps: Literal "\psn@sp\space startTexFig\space"}
\special{dvitps: Include1 "##1"}
\special{dvitps: Literal "endTexFig\space"}
}}
\def\psforDVIALW{
\def\PSspeci@l##1##2{
\special{language "PS"
literal "##2 1000 div dup scale"
include "##1"}}}
\def\psonlyboxes{
\def\PSspeci@l##1##2{%
\at(0cm;0cm){\boxit{\vbox to\drawinght
  {\vss
  \hbox to\drawingwd{\at(0cm;0cm){\hbox{(##1)}}\hss}
  }}}
}%
}
\def\psloc@lerr#1{%
\let\savedPSspeci@l=\PSspeci@l%
\def\PSspeci@l##1##2{%
\at(0cm;0cm){\boxit{\vbox to\drawinght
  {\vss
  \hbox to\drawingwd{\at(0cm;0cm){\hbox{(##1) #1}}\hss}
  }}}
\let\PSspeci@l=\savedPSspeci@l
}%
}
%
%
\newread\pst@mpin
\newdimen\drawinght\newdimen\drawingwd
\newdimen\psxoffset\newdimen\psyoffset
\newbox\drawingBox
\newif\ifNotB@undingBox
\newhelp\PShelp{Proceed: you'll have a 5cm square blank box instead of
your graphics (Jean Orloff).}
\def\@mpty{}
\def\s@tsize#1 #2 #3 #4\@ndsize{
  \def\psllx{#1}\def\pslly{#2}%
  \def\psurx{#3}\def\psury{#4}
  \ifx\psurx\@mpty\NotB@undingBoxtrue
  \else
    \drawinght=#4bp\advance\drawinght by-#2bp
    \drawingwd=#3bp\advance\drawingwd by-#1bp
  \fi
  }
\def\sc@nline#1:#2\@ndline{\edef\p@rameter{#1}\edef\v@lue{#2}}
\def\g@bblefirstblank#1#2:{\ifx#1 \else#1\fi#2}
\def\psm@keother#1{\catcode`#112\relax}
\def\execute#1{#1}
{\catcode`\%=12
\xdef\B@undingBox{
}   
\def\ReadPSize#1{
 \edef\PSfilename{#1}
 \openin\pst@mpin=#1\relax
 \ifeof\pst@mpin \errhelp=\PShelp
   \errmessage{I haven't found your postscript file (\PSfilename)}
   \psloc@lerr{was not found}
   \s@tsize 0 0 142 142\@ndsize
   \closein\pst@mpin
 \else
   \immediate\write\psbj@inaux{#1,}
   \loop
     \executeinspecs{\catcode`\ =10\global\read\pst@mpin to\n@xtline}
     \ifeof\pst@mpin
       \errhelp=\PShelp
       \errmessage{(\PSfilename) is not an Encapsulated PostScript File:
           I could not find any \B@undingBox: line.}
       \edef\v@lue{0 0 142 142:}
       \psloc@lerr{is not an EPSFile}
       \NotB@undingBoxfalse
     \else
       \expandafter\sc@nline\n@xtline:\@ndline
       \ifx\p@rameter\B@undingBox\NotB@undingBoxfalse
         \edef\t@mp{%
           \expandafter\g@bblefirstblank\v@lue\space\space\space}
         \expandafter\s@tsize\t@mp\@ndsize
       \else\NotB@undingBoxtrue
       \fi
     \fi
   \ifNotB@undingBox\repeat
   \closein\pst@mpin
 \fi
\message{#1}
}
%
%
\newcount\xscale \newcount\yscale \newdimen\pscm\pscm=1cm
\newdimen\d@mx \newdimen\d@my
\let\ps@nnotation=\relax
\def\psboxto(#1;#2)#3{\vbox{
   \ReadPSize{#3}
   \divide\drawingwd by 1000
   \divide\drawinght by 1000
   \d@mx=#1
   \ifdim\d@mx=0pt\xscale=1000
         \else \xscale=\d@mx \divide \xscale by \drawingwd\fi
   \d@my=#2
   \ifdim\d@my=0pt\yscale=1000
         \else \yscale=\d@my \divide \yscale by \drawinght\fi
   \ifnum\yscale=1000
         \else\ifnum\xscale=1000\xscale=\yscale
                    \else\ifnum\yscale<\xscale\xscale=\yscale\fi
              \fi
   \fi
   \divide \psxoffset by 1000\multiply\psxoffset by \xscale
   \divide \psyoffset by 1000\multiply\psyoffset by \xscale
   \global\divide\pscm by 1000
   \global\multiply\pscm by\xscale
   \multiply\drawingwd by\xscale \multiply\drawinght by\xscale
   \ifdim\d@mx=0pt\d@mx=\drawingwd\fi
   \ifdim\d@my=0pt\d@my=\drawinght\fi
   \message{scaled \the\xscale}
 \hbox to\d@mx{\hss\vbox to\d@my{\vss
   \global\setbox\drawingBox=\hbox to 0pt{\kern\psxoffset\vbox to 0pt{
      \kern-\psyoffset
      \PSspeci@l{\PSfilename}{\the\xscale}
      \vss}\hss\ps@nnotation}
   \global\ht\drawingBox=\the\drawinght
   \global\wd\drawingBox=\the\drawingwd
   \baselineskip=0pt
   \copy\drawingBox
 \vss}\hss}
  \global\psxoffset=0pt
  \global\psyoffset=0pt
  \global\pscm=1cm
  \global\drawingwd=\drawingwd
  \global\drawinght=\drawinght
}}
%
%
\def\psboxscaled#1#2{\vbox{
  \ReadPSize{#2}
  \xscale=#1
  \message{scaled \the\xscale}
  \divide\drawingwd by 1000\multiply\drawingwd by\xscale
  \divide\drawinght by 1000\multiply\drawinght by\xscale
  \divide \psxoffset by 1000\multiply\psxoffset by \xscale
  \divide \psyoffset by 1000\multiply\psyoffset by \xscale
  \global\divide\pscm by 1000
  \global\multiply\pscm by\xscale
  \global\setbox\drawingBox=\hbox to 0pt{\kern\psxoffset\vbox to 0pt{
     \kern-\psyoffset
     \PSspeci@l{\PSfilename}{\the\xscale}
     \vss}\hss\ps@nnotation}
  \global\ht\drawingBox=\the\drawinght
  \global\wd\drawingBox=\the\drawingwd
  \baselineskip=0pt
  \copy\drawingBox
  \global\psxoffset=0pt
  \global\psyoffset=0pt
  \global\pscm=1cm
  \global\drawingwd=\drawingwd
  \global\drawinght=\drawinght
}}
%
\def\psbox#1{\psboxscaled{1000}{#1}}
%
%
%
\newif\ifn@teof\n@teoftrue
\newif\ifc@ntrolline
\newif\ifmatch
\newread\j@insplitin
\newwrite\j@insplitout
\newwrite\psbj@inaux
\immediate\openout\psbj@inaux=psbjoin.aux
\immediate\write\psbj@inaux{\string\joinfiles}
\immediate\write\psbj@inaux{\jobname,}
%
%
\immediate\let\oldinput=\input
\def\input#1 {
 \immediate\write\psbj@inaux{#1,}
 \oldinput #1 }
\def\empty{}
\def\setmatchif#1\contains#2{
  \def\match##1#2##2\endmatch{
    \def\tmp{##2}
    \ifx\empty\tmp
      \matchfalse
    \else
      \matchtrue
    \fi}
  \match#1#2\endmatch}
\def\warnopenout#1#2{
 \setmatchif{TrashMe,psbjoin.aux,psbjoin.all}\contains{#2}
 \ifmatch
 \else
   \immediate\openin\pst@mpin=#2
   \ifeof\pst@mpin
     \else
     \errhelp{If the content of this file is so precious to you, abort (ie
press x or e) and rename it before retrying.}
     \errmessage{I'm just about to replace your file named #2}
   \fi
   \immediate\closein\pst@mpin
 \fi
 \message{#2}
 \immediate\openout#1=#2}
{
\catcode`\%=12
\gdef\splitfile#1 {
 \immediate\openin\j@insplitin=#1
 \message{Splitting file #1 into:}
 \warnopenout\j@insplitout{TrashMe}
 \loop
   \ifeof
     \j@insplitin\immediate\closein\j@insplitin\n@teoffalse
   \else
     \n@teoftrue
     \executeinspecs{\global\read\j@insplitin to\spl@tinline\expandafter
       \ch@ckbeginnewfile\spl@tinline
     \ifc@ntrolline
     \else
       \toks0=\expandafter{\spl@tinline}
       \immediate\write\j@insplitout{\the\toks0}
     \fi
   \fi
 \ifn@teof\repeat
 \immediate\closeout\j@insplitout}
\gdef\ch@ckbeginnewfile#1
 \def\t@mp{#1}
 \ifx\empty\t@mp
   \def\t@mp{#3}
   \ifx\empty\t@mp
     \global\c@ntrollinefalse
   \else
     \immediate\closeout\j@insplitout
     \warnopenout\j@insplitout{#2}
     \global\c@ntrollinetrue
   \fi
 \else
   \global\c@ntrollinefalse
 \fi}
\gdef\joinfiles#1\into#2 {
 \message{Joining following files into}
 \warnopenout\j@insplitout{#2}
 \message{:}
 {
 \edef\w@##1{\immediate\write\j@insplitout{##1}}
 \w@{
 \w@{
 \w@{
 \w@{
 \w@{
 \w@{
 \w@{
 \w@{
 \w@{\string\input\space psbox.tex}
 \w@{\string\splitfile{\string\jobname}}
 }
 \tre@tfilelist#1, \endtre@t
 \immediate\closeout\j@insplitout}
\gdef\tre@tfilelist#1, #2\endtre@t{
 \def\t@mp{#1}
 \ifx\empty\t@mp
   \else
   \llj@in{#1}
   \tre@tfilelist#2, \endtre@t
 \fi}
\gdef\llj@in#1{
 \immediate\openin\j@insplitin=#1
 \ifeof\j@insplitin
   \errmessage{I couldn't find file #1.}
   \else
   \message{#1}
   \toks0={
   \immediate\write\j@insplitout{\the\toks0}
   \executeinspecs{\global\read\j@insplitin to\oldj@ininline}
   \loop
     \ifeof\j@insplitin\immediate\closein\j@insplitin\n@teoffalse
       \else\n@teoftrue
       \executeinspecs{\global\read\j@insplitin to\j@ininline}
       \toks0=\expandafter{\oldj@ininline}
       \let\oldj@ininline=\j@ininline
       \immediate\write\j@insplitout{\the\toks0}
     \fi
   \ifn@teof
   \repeat
   \immediate\closein\j@insplitin
 \fi}
}
\def\autojoin{
 \immediate\write\psbj@inaux{\string\into\space psbjoin.all}
 \immediate\closeout\psbj@inaux
 \input psbjoin.aux
}
%
%
%
%
\def\centinsert#1{\midinsert\line{\hss#1\hss}\endinsert}
\def\psannotate#1#2{\def\ps@nnotation{#2\global\let\ps@nnotation=\relax}#1}
\def\pscaption#1#2{\vbox{
   \setbox\drawingBox=#1
   \copy\drawingBox
   \vskip\baselineskip
   \vbox{\hsize=\wd\drawingBox\setbox0=\hbox{#2}
     \ifdim\wd0>\hsize
       \noindent\unhbox0\tolerance=5000
    \else\centerline{\box0}
    \fi
}}}
\def\psfig#1#2#3{\pscaption{\psannotate{#1}{#2}}{#3}}
\def\psfigurebox#1#2#3{\pscaption{\psannotate{\psbox{#1}}{#2}}{#3}}
%
\def\at(#1;#2)#3{\setbox0=\hbox{#3}\ht0=0pt\dp0=0pt
  \rlap{\kern#1\vbox to0pt{\kern-#2\box0\vss}}}
%
\newdimen\gridht \newdimen\gridwd
\def\gridfill(#1;#2){
  \setbox0=\hbox to 1\pscm
  {\vrule height1\pscm width.4pt\leaders\hrule\hfill}
  \gridht=#1
  \divide\gridht by \ht0
  \multiply\gridht by \ht0
  \gridwd=#2
  \divide\gridwd by \wd0
  \multiply\gridwd by \wd0
  \advance \gridwd by \wd0
  \vbox to \gridht{\leaders\hbox to\gridwd{\leaders\box0\hfill}\vfill}}
%
\def\fillinggrid{\at(0cm;0cm){\vbox{
  \gridfill(\drawinght;\drawingwd)}}}
%
%
\def\textleftof#1:{
  \setbox1=#1
  \setbox0=\vbox\bgroup
    \advance\hsize by -\wd1 \advance\hsize by -2em}
\def\textrightof#1:{
  \setbox0=#1
  \setbox1=\vbox\bgroup
    \advance\hsize by -\wd0 \advance\hsize by -2em}
\def\endtext{
  \egroup
  \hbox to \hsize{\valign{\vfil##\vfil\cr%
\box0\cr%
\noalign{\hss}\box1\cr}}}
%
\def\frameit#1#2#3{\hbox{\vrule width#1\vbox{
  \hrule height#1\vskip#2\hbox{\hskip#2\vbox{#3}\hskip#2}%
        \vskip#2\hrule height#1}\vrule width#1}}
\def\boxit#1{\frameit{0.4pt}{0pt}{#1}}
\catcode`\@=12 
%
 \psfordvips   

\parskip=8pt plus 1pt minus 1pt
\baselineskip=14pt
\magnification\magstep1

\font \srm = cmr8 scaled \magstephalf

\font \brm = cmr10 scaled \magstep 2
\font \bbrm = cmr10 scaled \magstep 3
\font \bbf = cmbx10 scaled \magstep 2 
\def \vo{\vskip 5mm}
\def \vsm{\vskip 1cm}
\def \vsss{\vskip 2.5cm}
\def \hs {\hskip 0.5cm}
\def \ve{\vfill\eject}
\def \ce{\centerline}
\def \di{\displaystyle}
\def \d{\partial}
\def \artanh {\,\text{\rm artanh}\,}
\def \out{\text{\rm out}}
\def \odd{\text{\rm odd}}
\def \const{\text{\rm const}}
        \def \G{\Gamma}
        \def \g{\gamma}
        \def \a{\alpha}
        \def \b{\beta}
        \def \de{\delta}
        \def \De{\Delta}
        \def \ep{\varepsilon}
        \def \kappa{\varkappa}
        \def \la{\lambda}
        \def \La{\Lambda}
        \def \r{\rho}
        \def \t{\tau}
        \def \th{\theta}
        \def \z{\zeta}
        \def \Sg{\Sigma}
        \def \sg{\sigma}
        \def \Om{\Omega}
        \def \U{\Cal Q}
        \def \om{\omega}
        \def \Var{\text{\rm Var}\;}
        \def \tr{\,\text{\rm tr}\,}
        \def \sign{\,\text{\rm sign}\,}
        \def \f{\varphi}
        \def \N{\Bbb N}
        \def \Q{\Bbb Q}
        \def \R{\Bbb R}
        \def \C{\Bbb C}
        \def \T{\Bbb T}
        \def \Z{\Bbb Z}
        \def \rf{\root 4 \of}
        \def \Ai{\,\text{\rm Ai}\,}
        \def \iff{\quad\text{\rm if}\quad}
        \def \ON{O(N^{-2})}
        \def \iz{\int_{z_1}^{z_2}}
        \def \ix{\int_{x_1}^{x_2}}
        \def \A{\text{\rm A}}
        \def \E{\text{\rm E}\,}
        \def \lacr{\la_{\text{\rm cr}}} 
        \def \ov{\overline}     
\null\vskip 1cm 

\rightline {{\it Journ. Statist. Phys.} {\bf 88} (1997), 
269-305}

\vskip 1cm

\ce{\bbrm Correlations between Zeros of}  \vskip 0.3cm
\ce{\bbrm a Random Polynomial} \vskip  1cm

\ce{\brm Pavel Bleher${}^1$ and Xiaojun Di${}^1$}

\vskip 1cm 

\ce{\it Revised version}

\vfootnote {} {${}^1$Department of Mathematical Sciences, 
Indiana University -- Purdue University at Indianapolis, 402
N. Blackford Street, Indianapolis, IN 46202, USA. \hfill\break E-mail:
bleher\@math.iupui.edu, xdi\@math.iupui.edu.}







\vskip 2cm

{\bf Abstract.} We obtain exact analytical expressions for
correlations between real zeros of
the Kac random polynomial. We show that the zeros in the interval
$(-1,1)$ are 
asymptotically independent of the zeros outside of this interval, and 
that the straightened zeros have the same limit translation invariant 
correlations. Then we calculate the correlations
between the straightened zeros of the SO(2) random 
polynomial.

\vskip 1cm 





\vfill\eject

\beginsection 1. Introduction \par

Let $f_n(t)$ be a real random polynomial of degree $n$,
$$
f_n(t)=c_0+c_1t+\dots+c_nt^n,
\eqno(1.1)
$$
where $c_0,c_1,\dots,c_n$ are independent real random variables.
Distribution of zeros for various classes of random polynomials is
studied in the classical
papers by Bloch and 
Polya [BP], Littlewood and Offord [LO], Erd\"os and Offord [EO],
Erd\"os and Tur\'an [ET], and 
Kac [K1--K3].
We will  assume that the coefficients $c_0,c_1,\dots,c_n$ are normally
distributed with  
$$
\E c_j=0, \qquad \E c_j^2=\sigma_j^2.
\eqno(1.2)
$$
In the case when 
$$
\sg_j^2=1,
$$
  $f_n(t)$ is the Kac random polynomial. Another interesting case is
when 
$$
\sigma_j^2=
\pmatrix
n\\ j
\endpmatrix\,.
$$
As is pointed out by Edelman and Kostlan [EK], ``this particular
random 
polynomial is probably the more natural definition of a random
polynomial''. We call this polynomial the SO(2) random polynomial
because its $m$-point joint probability distribution of zeros is
SO(2)-invariant for all $m$ (see section 5 below). The SO(2) random
polynomial can 
be viewed as the Majorana spin state [Maj] with real random
coefficients, and 
it models a chaotic spin wavefunction in the Majorana
representation. See the papers by Leboeuf 
[Leb1, Leb2], 
Leboeuf and Shukla [LS], Bogomolny, Bohigas, and Leboeuf [BBL2], and 
Hannay [Han], where the SU(2) and some other random polynomials are
introduced and studied, 
that represent the Majorana spin states with complex  random
coefficients.  

Let $\{\t_1,\dots, \t_k\}$ be the set of real zeros of $f_n(t)$.
Consider the distribution function of the real zeros,
$$
P_n(t)=\E\,\#\{j\: \t_j\le t\},
$$
where the mathematical expectation is taken with respect to the joint 
distribution of the coefficients $c_0,\dots,c_n$. Let  
$$
p_n(t)=P'_n(t)
$$ 
be the density function.
By the Kac formula (see, e.g., [K3]),
$$
p_n(t)={\sqrt {A_n(t)C_n(t)-B_n^2(t)}\over \pi\,A_n(t)}.
\eqno (1.3)
$$
where
$$
\eqalign{
A_n(t)&=\sum_{j=0}^n \sg_j^2 t^{2j}\,,\cr  B_n(t)&=\sum_{j=1}^n
j\sg_j^2 t^{2j-1}={A'_n(t)\over 2}\,,\cr   
C_n(t)&=\sum_{j=1}^n j^2\sg_j^2 t^{2j-2}={A_n''(t)\over
4}+{A'_n(t)\over4t}\,.\cr}  
\eqno (1.4)
$$
The derivation of (1.3) by Kac is rather complex. A short proof of
(1.3) is given in the paper [EK] by Edelman and Kostlan. See also the
papers by Hannay [Han] and Mesincescu, Bessis, Fournier, Mantica, and
Aaron [M-A], and section 2 below. The formula (1.3) implies that for
the Kac random polynomial, 
$$
\lim_{n\to\infty} p_n(t)=p(t)={1\over \pi|1-t^2|},\qquad t\not=\pm1,
\eqno (1.5)
$$
and
$$
p_n(\pm 1)={1\over \pi}\left[{n(n+2)\over 12}\right]^{1/2}
$$  
(see [K3], [BS], and [EK]). The limiting density $p(t)$ is not
integrable at $\pm 1$, and this means that the zeros are mostly
located near 
$\pm 1$. Observe, in addition, that $p_n(t)$ is an even function of
$t$, and the distribution $p_n(t)dt$ is invariant with respect to the
transformation $t\to 1/t$. Kac [K1] proves that the expected number of 
real zeros has the asymptotics
$$
N_n=\int_{-\infty}^\infty p_n(t)\,dt
=(2/\pi)\log n+O(1).
$$
Kac [K2],
Erd\"os and Offord [EO], Stevens [Ste], 
Ibragimov and Maslova [IM], Logan and Shepp [LS], Edelman and
Kostlan [EK], and others extend
this asymptotics to various classes  of the random coefficients $\{
c_j\}$.  
Maslova [Mas1] evaluates the variance of the number of real zeros as
$$
\Var \#\{j\: f_n(\t_j)=0\}={4\over \pi}\left(1-{2\over \pi}\right)\ln
n (1+o(1)),\qquad n\to\infty,
$$
and she proves the central limit theorem for the number of real zeros
(see [Mas2]), for a class of distributions of the random coefficients
$\{c_j\}$. 

In this paper we are interested in correlations between the
zeros $\t_j$ of the Kac random polynomial. Let us consider first
the zeros in the interval $(-1,1)$.
Define straightening  of $\t_j$ as
$$
\z_j=P(\t_j), \qquad P(t)=\int_0^t p(u)\,du.
$$
In the limit when $n\to\infty$, the straightened zeros $\z_j$ are
uniformly distributed on the real line, so that
$$
\lim_{n\to\infty} \E\,\#\,\{j\: a<\z_j\le b\}=b-a.
\eqno (1.6)
$$
From (1.5) we get that
$$
P(t)=\int_0^t {du \over \pi(1-u^2)}={1\over
2\pi}\ln\left|{1+t\over 
1-t}\right| ={1\over \pi}\artanh t,
$$
hence
$$
\z_j={1\over \pi}\artanh \t_j.
\eqno (1.7)
$$
Let $p_{mn}(s_1,\dots, s_m)$ be the joint probability distribution
density of the straightened zeros $\z_j$,
$$
p_{nm}(s_1,\dots, s_m)=\lim_{\De s_1,\dots,\De s_m\to 0} 
{\Pr\,\{ \exists\, \z_{j_1}\in [s_1,s_1+\De s_1],\dots,
\exists\, \z_{j_m}\in [s_m,s_m+\De s_m]\}\over |\De s_1\dots \De
s_m|}. 
\eqno (1.8)
$$
It coincides with the correlation function
$$
k_{nm}(s_1,\dots, s_m)= \lim_{\De s_1,\dots,\De s_m\to 0}
{\E\, \bigl[\xi_n(s_1,s_1+\De s_1)\dots \xi_n(s_m,s_m+\De s_m)\bigr]
\over |\De s_1\dots \De s_m|}.
\eqno (1.9)
$$
where
$$
\xi_n(a,b)=\#\,\{j: a< \z_j\le b\}.
$$
We assume in (1.8) and (1.9) that $s_i\not=s_j$ for all $i\not=j$.
Our aim is to find the  limit correlation functions
$$
k_m(s_1,\dots, s_m)=\lim_{n\to\infty} k_{nm}(s_1,\dots, s_m).
\eqno (1.10)
$$
We prove the following results.

{\bf Theorem 1.1.} {\it 
The limit two-point correlation function $k_2(s_1,s_2)$ of
the straightened zeros $\z_j=\pi^{-1}\artanh \t_j$ of the Kac
random polynomial is equal to
$$
k_2(s_1,s_2)=\tanh^2{\pi (s_1-s_2)}+{|\sinh{\pi (s_1-s_2)}|\over
\cosh^2{\pi(s_1-s_2)}}\arcsin{1\over \cosh{\pi (s_1-s_2)}}   
\eqno(1.11)
$$}

Observe that $k_2(s_1,s_2)$ depends only on $s_1-s_2$, and it has the
following asymptotics:
$$
\eqalign{
k_2(s_1,s_2)&={\pi^2\over
2}|s_1-s_2|+O(|s_1-s_2|^2),\qquad|s_1-s_2|\to 0, \cr 
k_2(s_1,s_2)&=1-{16\over 3}\,e^{-4\pi|s_1-s_2|}+O(e^{-6\pi|s_1-s_2|}),
\qquad |s_1-s_2| \to \infty. \cr} 
$$
The graph of $k_2(0,s)$ is given in Fig. 1.

\centinsert{\pscaption{\psboxto (6in;2.4in){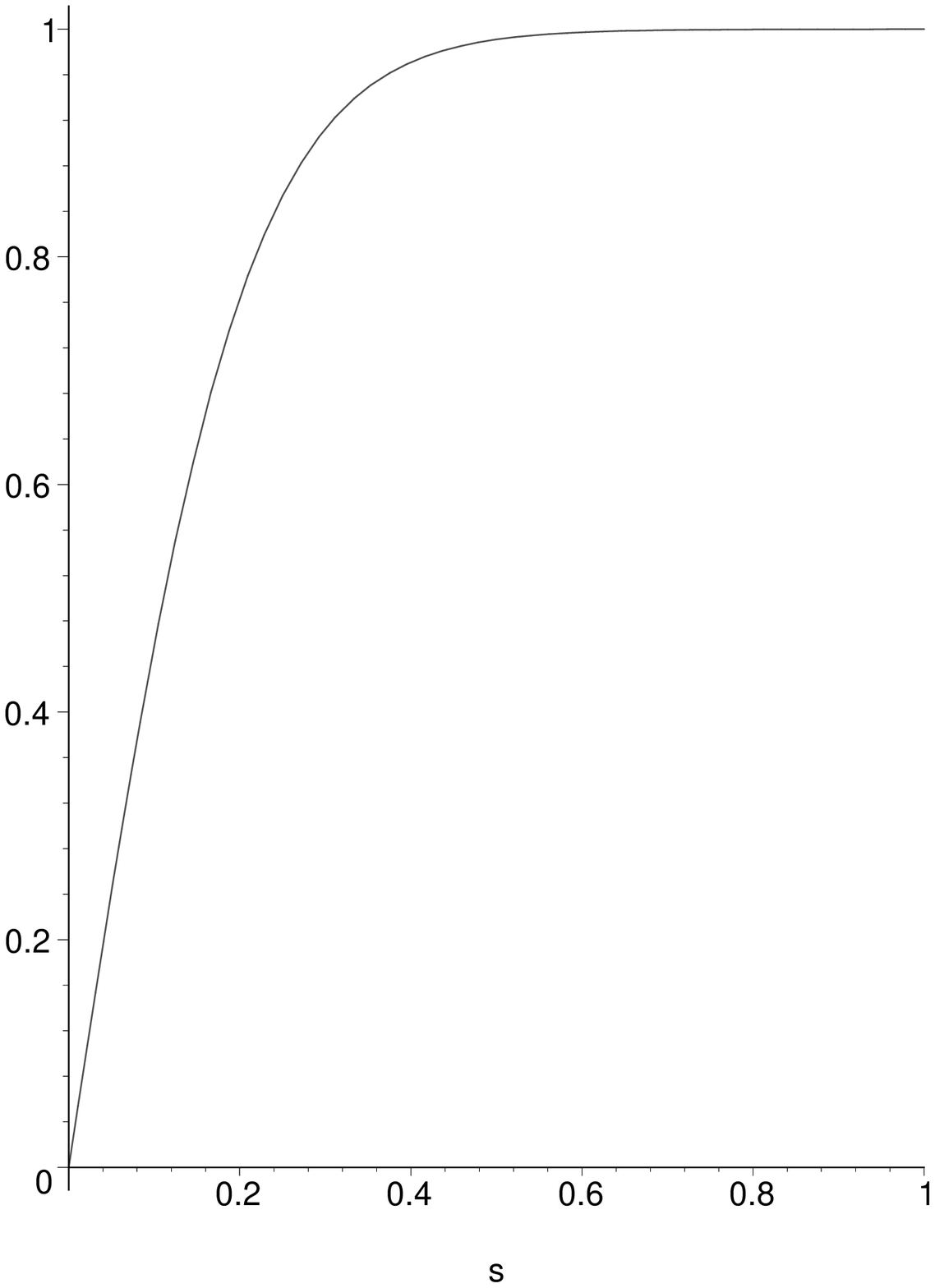}}
{\srm Fig 1:  The two-point correlation function of straightened zeros of
the Kac random polynomial. }}

{\bf Theorem 1.2.} {\it 
The limit $m$-point correlation function $k_m(s_1,\dots,s_m)$ of the
straightened zeros $\z_j=\pi^{-1}\artanh \t_j$ is equal to
$$
k_m(s_1,\dots,s_m)
= 2^{-m}\int_{-\infty}^{\infty}\dots\int_{-\infty}^{\infty}
|y_1\dots y_m|e^{-{1\over 2}(Y\Gamma_m,Y)}\,dy_1\dots dy_m,
\eqno(1.12)
$$
where $Y=(y_1,\dots,y_m)$ 
and the matrix $\Gamma_m$ is defined as
$$
\Gamma_m=\left({1\over \cosh\pi (s_i-s_j)}\right)_{i,j=1}^m
\eqno(1.13)
$$
In particular, $k_m(s_1,\dots,s_m)$ depends only on the differences of
$s_1,\dots,s_m$, hence it is translation  invariant.} 

The proof of Theorems 1.1 and 1.2 is given in sections 2, 3 and 4
below. 
It is based on computation of the determinant of some matrices which
consist of $2\times 2$ blocks. This computation is of independent
interest. The basic  example is the matrix 
$$ 
\Delta_m=\pmatrix
\Delta_{11}&\Delta_{12}& \dots &\Delta_{1m} \cr
\Delta_{21}&\Delta_{22}& \dots &\Delta_{2m} \cr
\dots&\dots&\dots&\dots \cr
\Delta_{m1}&\Delta_{m2}& \dots &\Delta_{mm} \cr
\endpmatrix
\eqno(1.14)
$$
where
$$
\Delta_{ij}=\pmatrix
\di {1\over{1-t_it_j}}&\di {t_i\over{(1-t_it_j)^2}} \cr
\di {t_j\over{(1-t_it_j)^2}}&\di {{1+t_it_j}\over{(1-t_it_j)^3}}\cr
\endpmatrix,
\qquad i,j=1,\dots,m.
\eqno(1.15)
$$  
We prove in section 4 that
$$
\det\Delta_m={{\prod_{1\le i<j\le m}(t_i-t_j)^8}\over
{\prod_{i=1}^m(1-t_i^2)^4\prod_{1\le i<j\le m}(1-t_it_j)^8}} 
\eqno(1.16)
$$
It is interesting to note that determinants of matrices consisting of
$2\times 2$ blocks appear also in the theory of random matrices (see,
e.g., [Dys] and [Meh]), statistical physics, and other
applications.   

Consider now zeros $\t_j$ with $|\t_j|>1$. Define straightening of
$\t_j$ as
$$
\z_j=P(\t_j),\qquad P(t)=
\left\{
\eqalign{
&\int_t^{-\infty}p(u)\,du\iff t<-1,\cr
&\int_t^\infty p(u)\,du\iff t>1.\cr}
\right.
\eqno (1.17)
$$
In the limit when $n\to\infty$, the straightened zeros $\z_j$ are
uniformly 
distributed on the real line. From (1.5)
$$
P(t)={1\over
2\pi}\ln\left|{1+t\over 
1-t}\right| ={1\over \pi}\artanh t^{-1},
\eqno (1.18)
$$
so that
$$
\z_j=\pi^{-1}\artanh \t_j^{-1}.
\eqno (1.19)
$$
Denote by $k_{nm}^{\out}(s_1,\dots,s_m)$ the correlation function of
the straightened zeros $\z_j$ with $|\t_j|>1$.

{\bf Theorem 1.3.}
$$
k_{nm}^{\out}(s_1,\dots,s_m)=k_{nm}(s_1,\dots,s_m).
\eqno (1.20)
$$

In other words, the correlation functions of the straightened zeros
outside of the interval $(-1,1)$ coincide with those inside of the
interval. Finally, let us consider correlation between zeros
inside of the interval $(-1,1)$ and outside of this interval. Let
$K_{nm}(t_1,\dots,t_m)$ be the correlation function of the zeros
$\t_j$ (without straightening).

{\bf Theorem 1.4.} {\it Assume that $|t_1|,\dots,|t_l|<1$ and
$|t_{l+1}|,\dots, |t_m|>1$. Then the limit
$$
\lim_{n\to\infty}K_{nm}(t_1,\dots,t_m)
=K_{m}(t_1,\dots,t_m)
\eqno (1.21)
$$
exists and
$$
K_{m}(t_1,\dots,t_m)=K_l(t_1,\dots,t_l)\, K_{m-l}(t_{l+1},\dots t_m).
\eqno (1.22)
$$}

This means that the zeros inside and outside of the interval $(-1,1)$
are asymptotically independent. Observe that
$$
k_{m}(s_1,\dots,s_m)=
\left[{K_{m}(t_1,\dots,t_m)\over p(t_1)\dots
p(t_m)}\right]_{t_1=P^{-1}(s_1),\dots,t_m=P^{-1}(s_m)} \,,
\eqno (1.23)
$$
provided that either all $|t_j|<1$ or all $|t_j|>1$ (cf. the formula
(2.14) below). 
Proof of Theorems 1.3 and 1.4 is given in the end of section 4.

In sections 5 and 6 we investigate correlation functions of
real zeros of the SO(2) random polynomial. 
     
\beginsection 2. General Formulae \par

Let 
$$
f_n(t)=\sum_{j=0}^n c_jt^j, \eqno (2.1)
$$
be a polynomial whose coefficients $c_j$ are random variables with
an absolutely continuous joint distribution. 
Let 
$$
\xi_n{(a,b)}=\#\{\t_k: a<\t_k\le b,\; f_n(\t_k)=0\}
\eqno(2.2)
$$
be the number of real roots of $f_n(t)$ between $a$ and $b$, and
let $p_n(t)$ be the density of real
zeros $t_k$ of $f_n(t)$, so that 
$$
\E\,\xi_n(a,b)=\int_a^b p_n(t)\,dt.\eqno (2.3)
$$
It is not difficult to show that
$$
p_n(t)=\int_{-\infty}^\infty |y|\,D_n(0,y;t)\,dy,
\eqno (2.4)
$$
where $D_n(x,y;t)$ is a joint distribution density of $f_n(t)$ and
$f'_n(t)$,  
$$  
\Pr \;\{\;a<f_n(t)\le b;\; c<f'_n(t)\le d\;\}
=\int_a^b\int_c^d D_n(x,y;t)\,dxdy. 
\eqno (2.5)
$$
Indeed, if $f_n'(t)=y$ then asymptotically as $\De t\to 0$,
the function $f_n(t)$ has a zero in the interval $[t,t+\De t]$ if
$f_n(t)$ is in the interval $[0,-y\De t]$, and this gives (2.5).   
Similarly, the $m$-point correlation function $K_{nm}(t_1,\dots,t_m)$
for pairwise different $t_1,\dots, t_m$ is equal to
$$
K_{nm}(t_1,\dots,t_m)=\int_{-\infty}^\infty\dots \int_{-\infty}^\infty
|y_1\dots y_m| D_{nm}(0,y_1,\dots, 0,y_m;t_1,\dots,t_m)dy_1\dots dy_m,
\eqno (2.6)
$$
where $D_{nm}(x_1,y_1,\dots,x_m,y_m;t_1,\dots,t_m)$ is a joint
distribution density of  the vector
$$
F_n=(f_n(t_1),f'_n(t_1),\dots,f_n(t_m),f'_n(t_m)),
$$
 so that
$$\eqalign{
\Pr\,&\{ a_1<f_n(t_1)\le b_1;\; c_1< f'_n(t_1)\le d_1;\;\dots\;;\;a_m<
f_n(t_m)\le b_m;\; c_m<f'_n(t_m)\le d_m\}\cr 
&=\int_{a_1}^{b_1}\int_{c_1}^{d_1}\dots
\int_{a_m}^{b_m}\int_{c_m}^{d_m}
D_{nm}(x_1,y_1,\dots,x_m,y_m;t_1,\dots,t_m)\,dx_1dy_1\dots dx_mdy_m.\cr}
\eqno (2.7)
$$
If $\{ c_j\}$ are independent random variables with $\Var c_j>0$ then
the covariance matrix of the vector $F_n$ is positive, provided that
$n\ge 2m-1$ (see Appendix B at the end of the paper).
Similar formulae are derived for the correlation functions of complex
zeros of random polynomials with complex and real coefficients (see
[Han] and [M-A]). 

Observe that
$$ \E\,\prod_{j=1}^m\xi_n(a_j,b_j)=\int_{a_1}^{b_1}\dots
\int_{a_m}^{b_m}K_{nm}(t_1,\dots,t_m)\,dt_1\dots dt_m,  
\eqno (2.8) 
$$
provided that $(a_1,b_1),\dots,(a_m,b_m)$ are pairwise disjoint, and
$$
p_n(t)=K_{n1}(t),\qquad \E(\xi_n(a,b))=\int_a^bK_{n1}(t)dt.
\eqno(2.9)
$$
For the general case, when $(a_1,b_1),\dots,(a_m,b_m)$ may
intersect, we have the following extension of (2.8):
$$
\E\prod_{j=1}^m\xi_n(a_j,b_j)=\sum_{(A_1,\dots,A_l)}
\prod_{j=1}^l \left(\int_{\bigcap\limits_{i\in A_j}(a_i,b_i)}dt_j
\right)K_{nl}(t_1,\dots,t_l), 
\eqno(2.10)
$$
where the sum is taken over all possible partitions $(A_1,\dots,A_l)$
of $\{1,\dots,m\}$, such that 
$$\eqalign{
&A_i\cap A_j=\emptyset,\qquad i\not=j,\cr
&A_1\cup \dots \cup A_l=\{1,\dots,m\},\cr
&|A_i|\geq1 \qquad i=1,\dots,l.\cr}
\eqno(2.11)
$$
In particular, when $m=2$ we have
$$
\E\,[\xi_n(a_1,b_1)\xi_n(a_2,b_2)]=\int_{a_1}^{b_1}\int_{a_2}^{b_2}
K_{n2}(t_1,t_2)dt_1dt_2 
\eqno(2.12)
$$
if $(a_1,b_1)\cap(a_2,b_2)=\emptyset$, and
$$
\E\,[\xi^2_n(a,b)]=\int_{a}^{b}p_n(t)dt+\int_{a}^{b}\int_{a}^{b}
K_{n2}(t_1,t_2)dt_1dt_2 
\eqno(2.13)
$$
From definition (1.9) of the $m$-point correlation function, it
follows 
that the $m$-point correlation function $k_{nm}(s_1,\dots,s_m)$ of the
straightened zeros $\z_j=P(\t_j)$ is related to the $m$-point
correlation function $K_{nm}(t_1,\dots,t_m)$ of the zeros $\t_j$ by
the formula
$$
k_{nm}(s_1,\dots,s_m)=
\left[{K_{nm}(t_1,\dots,t_m)\over p(t_1)\dots
p(t_m)}\right]_{t_1=P^{-1}(s_1),\dots,t_m=P^{-1}(s_m)} \,.
\eqno (2.14)
$$
Assume now that the coefficients $c_j$ are independent Gaussian
variables with zero mean and the variances $\sg^2_j$, $j=0,\dots, n$.
Then $D_{n1}(x,y;t)$ is a Gaussian distribution density with the
covariance matrix 
$$
\De=\pmatrix
\E\,f_n^2(t) & \E\, f_n(t)f'_n(t) \\
\E\, f_n(t)f'_n(t) & \E\, (f'_n(t))^2 
\endpmatrix=\pmatrix
A_n(t) & B_n(t) \\
B_n(t) & C_n(t) 
\endpmatrix\,,
\eqno (2.15)
$$
where $A_n(t)$, $B_n(t)$ and $C_n(t)$ are defined in (1.4),
and from (2.4) we get the Kac formula (1.3).

\beginsection 3. Two-Point Correlation Function for the Kac Polynomial
\par

   Let $f_n(t)=c_0+c_1t+\dots +c_nt^n$ be the Kac polynomial, so that
 $c_k, k=0,\dots,n$, are real independent Gaussian random
variables with  
$$
\E {c_k}=0, \qquad \E {c_k}^2=1.
\eqno(3.1)
$$
Consider the covariance matrix $\De_n$ of the Gaussian vector
$(f_n(t_1),f_n'(t_1),f_n(t_2),f_n'(t_2))$. From (3.1)
$$\eqalign{
&\E\, f_n(t_1)f_n(t_2)=\sum_{k=0}^n (t_1t_2)^k={1-(t_1t_2)^{n+1}\over
1-t_1t_2}\,, \cr
&\E\, f_n'(t_1)f_n(t_2)={\partial\over\partial t_1}\left[
{1-(t_1t_2)^{n+1}\over
1-t_1t_2}\right]\,,\cr
&\E\, f_n'(t_1)f_n'(t_2)={\partial^2\over\partial t_1\partial t_2}\left[
{1-(t_1t_2)^{n+1}\over
1-t_1t_2}\right]\,.\cr}
\eqno (3.2)
$$
Assume that $|t_1|,|t_2|<1$. Then from (3.2) we obtain that
$$
\lim_{n\to\infty}\De_n=\De,
\eqno (3.3)
$$
with 
$$
\Delta=\pmatrix
1\over{1-{t_1}^2}&{t_1\over{(1-{t_1}^2)^2}}&{1\over{1-t_1t_2}}
&{t_1\over{(1-t_1t_2)^2}} \cr
{t_1\over{(1-{t_1}^2)^2}}&{{1+{t_1}^2}\over{(1-{t_1}^2)^3}}
&{t_2\over{(1-t_1t_2)^2}}&{{1+t_1t_2}\over(1-t_1t_2)^3} \cr
{1\over{1-t_1t_2}}&{{1+t_1t_2}\over(1-t_1t_2)^3}&{1\over{1-{t_2}^2}}&{t_2\over{(1-{t_2}^2)^2}} \cr
{t_1\over{(1-t_1t_2)^2}}&{{1+t_1t_2}\over(1-t_1t_2)^3} &{t_2\over{(1-{t_2}^2)^2}}&{{1+{t_2}^2}\over{(1-{t_2}^2)^3}} \cr
\endpmatrix\,.
\eqno(3.4)
$$
We prove in the section 4 below that 
$$
\det\Delta={(t_1-t_2)^8\over(1-{t_1}^2)^4(1-{t_2}^2)^4(1-t_1t_2)^8}\,.
\eqno(3.5)
$$
Let $\Omega$ be the two-by-two matrix obtained by removing the first
and the third rows and columns from $\Delta^{-1}$. Then
$$
\Omega=\pmatrix
A&B \cr
B&C \cr
\endpmatrix
\eqno(3.6)
$$
where 
$$\eqalign{
A&=(1-t_1t_2)^4(1-{t_1}^2)^3 /(t_1-t_2)^4\cr
B&=(1-t_1t_2)^3(1-{t_1}^2)^2(1-{t_2}^2)^2/(t_1-t_2)^4 \cr
C&=(1-t_1t_2)^4(1-{t_2}^2)^3 /(t_1-t_2)^4 \cr}
\eqno(3.7)
$$
By (2.6), the correlation function $K_2(t_1,t_2)$ 
is equal to 
$$
K_2(t_1,t_2)={1\over{{4\pi}^2\sqrt{\det
\Delta}}}\int_{-\infty}^\infty\int_{-\infty}^\infty|y_1y_2|
e^{{-{1\over2}(Y\Omega,Y)}}dy_1dy_2
\eqno(3.8)
$$
where $Y=(y_1,y_2)$. Since 
$$\eqalign{
\int_{-\infty}^{\infty}\int_{-\infty}^{\infty}
&|y_1y_2|e^{{-{1\over 2}}(Ay_1^2+2By_1y_2+Cy_2^2)}\,dy_1dy_2\cr
&={4\over AC(1-\de^2)}
\left(1+{\de\over \sqrt{1-\de^2}}\arcsin \de\right),\qquad
\de={{B}\over\sqrt{AC}}\cr}
\eqno(3.9)
$$
(see Appendix A), we obtain that
$$\eqalign{
K_2(t_1,t_2)=&{(t_1-t_2)^2\over{\pi^2(1-t_1t_2)^2(1-{t_1}^2)(1-{t_2}^2)}} \cr
&+{|t_1-t_2|\over{\pi^2(1-t_1t_2)^2}\sqrt{(1-{t_1}^2)(1-{t_2}^2)}}
\arcsin{{\sqrt{(1-{t_1}^2)(1-{t_2}^2)}}\over{1-t_1t_2}}\cr} 
\eqno(3.10)
$$
Consider the correlation function $k_2(s_1,s_2)$ of the straightened
zeroes $\z_j=\pi^{-1}\artanh \t_j$. By (2.14),
$$
k_2(s_1,s_2)={K_2(t_1,t_2)\over
p(t_1)p(t_2)}\,,\qquad t_1=\tanh(\pi s_1),\; t_2=\tanh (\pi s_2).
\eqno (3.12)
$$
Since 
$$
p(t)={1\over \pi(1-t^2)},
$$
(see (1.5)), we obtain that
$$\eqalign{
k_2(s_1,s_2)&={(t_1-t_2)^2\over (1-t_1t_2)^2}
+{|t_1-t_2|\sqrt{(1-{t_1}^2)(1-{t_2}^2)}\over(1-t_1t_2)^2}
\,\arcsin{{\sqrt{(1-{t_1}^2)(1-{t_2}^2)}}\over{1-t_1t_2}}\cr
&=\tanh^2{\pi (s_1-s_2)}+{|\sinh{\pi (s_1-s_2)|}\over
\cosh^2{\pi(s_1-s_2)}}\arcsin{1\over \cosh{\pi (s_1-s_2)}}   \cr}
$$
Theorem 1.1 is proved.

\beginsection 4. Higher order correlation functions for the Kac
polynomial   \par 

Let $f_n(t)$ be the Kac polynomial, and let
$t_1,t_2,\dots,t_m$ be 
$m\ge 3$ distinct points in the interval $(-1,1)$.
Denote by $\De^{(n)}_{m}$ the covariance matrix of the Gaussian vector
$$
(f_n(t_1),f'_n(t_1),\dots,f_n(t_m),f'_n(t_m)),
$$
and by $\Delta_m$ the limit of $\De^{(n)}_{m}$ as $n\to \infty$,   
$$
\Delta_m=\lim_{n\to \infty}\De^{(n)}_{m}
\eqno(4.1)
$$
Then
$$
\De^{(n)}_{m}=\pmatrix
\De^{(n)}_{11}&\De^{(n)}_{12}&\dots &\De^{(n)}_{1m} \cr
\De^{(n)}_{21}&\De^{(n)}_{22}& \dots &\De^{(n)}_{2m} \cr
\dots&\dots&\dots&\dots \cr
\De^{(n)}_{m1}&\De^{(n)}_{m2}& \dots &\De^{(n)}_{mm} \cr
\endpmatrix
\eqno(4.2)
$$
where
$$
\De^{(n)}_{ij}=\pmatrix
\E{f_n(t_i)f_n(t_j)}&\E{f_n(t_i)f^{'}_n(t_j)} \cr
\E{f^{'}_n(t_i)f_n(t_j)}&\E{f'_n(t_i)f'_n(t_j)} \cr
\endpmatrix
\eqno(4.3)
$$
and by (3.2),
$$ 
\Delta_m=\pmatrix
\Delta_{11}&\Delta_{12}& \dots &\Delta_{1m} \cr
\Delta_{21}&\Delta_{22}& \dots &\Delta_{2m} \cr
\dots&\dots&\dots&\dots \cr
\Delta_{m1}&\Delta_{m2}& \dots &\Delta_{mm} \cr
\endpmatrix
\eqno(4.4)
$$
where
$$
\Delta_{ij}=\pmatrix
{1\over{1-t_it_j}}& {t_i\over{(1-t_it_j)^2}} \cr
 {t_j\over{(1-t_it_j)^2}}& {{1+t_it_j}\over{(1-t_it_j)^3}}\cr
\endpmatrix
\eqno(4.5)
$$
[cf. (3.4)].

If $\Omega_m$ denotes the $m\times m$ matrix obtained by removing all
the odd number rows and columns from $\Delta^{-1}_m$, then by (2.6),
the correlation function $K_m(t_1,\dots,t_m)$ is
equal to  
$$
K_{m}(t_1,\dots,t_m)={1\over{(2\pi)^m\sqrt{\det\Delta_m}}}
\int_{-\infty}^{\infty}\dots\int_{-\infty}^{\infty}
|y_1\dots y_m|e^{-{1\over 2}(Y\Omega_m,Y)}\,dy_1\dots dy_m
\eqno(4.6)
$$
where $Y=(y_1,\dots,y_m)$. We have the following extension of the
formula (3.6).   

{\bf Proposition 4.1} 
$$
\det\Delta_m={{\prod_{1\le i<j\le m}(t_i-t_j)^8}\over
{\prod_{i=1}^m(1-t_i^2)^4\prod_{1\le i<j\le m}(1-t_it_j)^8}} 
\eqno(4.7)
$$

The proof of Proposition 4.1 uses the following lemma.

{\bf Lemma 4.2} {\it Let $f_n(t)$ $(n\ge 3)$ be any random polynomial
and $t_1,\dots,t_m$ be any m real numbers. Let $\De^{(n)}_m$ be the
covariance matrix of the Gaussian random vector
$$
(f_n(t_1),f'_n(t_1),\dots,f_n(t_m),f'_n(t_m))
$$ 
which is defined in (4.2) and (4.3). Then   
$$
\det \De^{(n)}_m=P_n(t_1,\dots,t_m)\prod_{1 \le i<j \le m}(t_i-t_j)^8  
\eqno(4.8)
$$
where $P_n(t_1,\dots,t_m)$ is a polynomial.}

{\it Proof.} To simplify notation we drop the indices $m,n$ in the
matrix $\De ^{(n)}_m$.
We have 
$$
\De=(\De_{ij})_{i,j=1,\dots, m}
$$
where
$$
\De_{ij}=\pmatrix
\E{f_n(t_i)f_n(t_j)}&\E{f_n(t_i)f'_n(t_j)} \cr
\E{f'_n(t_i)f_n(t_j)}&\E{f'_n(t_i)f'_n(t_j)} \cr
\endpmatrix
\eqno(4.10)
$$
In the following discussion we consider linear transformations of the
matrix $\De$ which do not change its determinant. 
By substracting the first and second column of $\De_{i1}$ from the first
and second column of $\De_{ij}$, respectively, we get the matrix
$\De^{(1)}$ with the $2\times 2$ blocks
$$
\De_{ij}^{(1)}=\pmatrix
\E{f_n(t_i)(f_n(t_j)-f_n(t_1))}&\E{f_n(t_i)(f'_n(t_j)-f'_n(t_1))}\cr
\E{f'_n(t_i)(f_n(t_j)-f_n(t_1))}&\E{f'_n(t_i)(f'_n(t_j)-f'_n(t_1))}\cr
\endpmatrix
\eqno(4.11)
$$
Since $f_n(t)$ is a polynomial, we can take the factor $(t_j-t_1)$
out of the first two columns of the matrix $\De^{(1)}$, and this
proves 
that $\det \De$ is divisible by $(t_j-t_1)^2$. Repeating the same
operation on rows we get the factor $(t_j-t_1)^4$. How to get
$(t_j-t_1)^8$? To that end let us subtract the second column of the
matrix $\De^{(1)}_{i1}=\De_{i1}$ multiplied by $(t_j-t_1)$ from the
first 
column of the matrix $\De^{(1)}_{ij}$. This produces the matrix 
$$
\De_{ij}^{(2)}=\pmatrix
\E f_n(t_i)\left[f_n(t_j)
-f_n(t_1)-(t_j-t_1)f'_n(t_1)\right] &\E f_n(t_i)\left[f'_n(t_j)
-f'_n(t_1)\right]\cr
\E f'_n(t_i)\left[f_n(t_j)-f_n(t_1)-(t_j-t_1)f'_n(t_1)\right]
&\E f'_n(t_i)\left[f'_n(t_j)
-f'_n(t_1)\right]\cr
\endpmatrix
\eqno(4.12)
$$
Now we can take $(t_j-t_1)^2$ out of the first column and $(t_j-t_1)$
out of the second column of the matrix $\De^{(2)}$. Repeating the same
operations over the rows we get that $\det \De$ is divisible by
$(t_j-t_1)^6$. Finally, let us observe that by the Taylor formula
$$
f_n(t_j)-f_n(t_1)-(t_j-t_1)f_n(t_1)={(t_j-t_1)^2\over
2}f''_n(t_1)+O(|t_j-t_1|^3), 
$$
and
$$
f_n'(t_j)-f_n'(t_1)=(t_j-t_1)f''_n(t_1)+O(|t_j-t_1|^2),
$$
hence if we subtract the second column of the matrix $\De^{(2)}_{ij}$ 
multiplied by $(t_j-t_1)/2$ from its first column, the difference is
of the order of $|t_j-t_1|^3$, and we can take the factor
$(t_j-t_1)^3$ 
out of the first column and $(t_j-t_1)$ out of the second column. This 
gives the factor $(t_j-t_1)^4$. The same factor is taken out of the
raws, hence $\det\De$ is divisible by $(t_j-t_1)^8$. Similarly, it is
divisible by $(t_j-t_i)^8$ for all $i\not= j$, and hence it is
divisible by their product. Lemma 4.2 is proved.

{\it Proof of  Proposition 4.1}. By (4.1), we have
$$
\det \Delta_m=\lim_{n \to \infty}\det \De^{(n)}_m
\eqno(4.13)
$$
for all $t_1,\dots,t_m$ in the interval $(-1,1)$. In fact, the limit
(4.13) holds 
for all complex $t_1,\dots,t_m$ in the unit disk $\{|t|<1\}$, and the
convergence is uniform on every disk $\{|t|<r\}$ where $r<1$. 
Hence by Lemma 4.2, 
$$
\det \Delta_m=H(t_1,\dots,t_m)\prod_{1 \le i<j \le m}(t_i-t_j)^8
\eqno(4.14)
$$
where $H(t_1,\dots,t_m)$ is holomorphic in the unit disk.

Now, let us consider the expression of $\det\Delta_m$ in terms of
the matrix elements $\de_{ij}$ of  $\Delta_m$, that is  
$$ 
\det\Delta_m=\sum_{\sigma}\epsilon_{\sigma}
\delta_{1\sigma(1)}\dots\delta_{n\sigma(2m)}
\eqno(4.15)
$$
where $\sigma$ is a permutation of $\{1,\dots,2m\}$ and
$\epsilon_m=\pm 1$ depending on whether $\sigma$ is even or odd.
The common denominator of the sum in (4.15) is  
$$
\prod_{i=1}^m(1-t_i^2)^4\prod_{1 \le i<j \le m}(1-t_it_j)^8
\eqno(4.16)
$$
Therefore by (4.14),
$$
\det\Delta_m={{\prod_{1 \le i<j\le m}(t_i-t_j)^8
C(t_1,\dots,t_m)}\over {\prod_{i=1}^m(1-t_i^2)^4
\prod_{1 \le i<j \le m}(1-t_it_j)^8}}
\eqno(4.17)
$$
where $C(t_1,\dots,t_m)$ is a polynomial of $t_1,\dots,t_m$. Observe
that (4.17) holds for all points $t_1,\dots,t_m$ in the unit
disk,  and so it can be extended to the whole complex plane.  
We are going to show that $C(t_1,\dots,t_m)$ is a constant, and
moreover, that  
$$
C(t_1,\dots,t_m)= 1
\eqno(4.18)
$$
Let us look at the asymptotic behavior of $\det\Delta_m$ as
$t_1\to \infty$ while $t_2,\dots,t_m$ are fixed. To see it more
clearly, let us change $\Delta_m$ to $\Delta_m^{(1)}$ by substracting
the $(2i-1)$th column and row from $(2i)$th column and row,
respectively, for $i=1,\dots,m$. Then 
$$
\Delta_m^{(1)}=\pmatrix
\Delta_{11}^{(1)}&\Delta_{12}^{(1)}&\dots&\Delta_{1m}^{(1)} \cr
\Delta_{21}^{(1)}&\Delta_{22}^{(1)}&\dots&\Delta_{2m}^{(1)} \cr
\dots&\dots&\dots&\dots \cr
\Delta_{m1}^{(1)}&\Delta_{m2}^{(1)}&\dots&\Delta_{mm}^{(1)} \cr
\endpmatrix
\eqno(4.19)
$$
where 
$$
\Delta_{kk}^{(1)}=\pmatrix
{1\over{1-t_k^2}}&0 \cr
0&{1\over{(1-t_k^2)}^3} \cr
\endpmatrix
\eqno(4.20)
$$
for $k=1,\dots,m,$ and
$$
\Delta_{ij}^{(1)}=\pmatrix
{1\over{1-t_it_j}}&{{t_i-t_j}\over{(1-t_it_j)^2(1-t_j^2)}} \cr
{{t_i-t_j}\over{(1-t_it_j)^2(1-t_j^2)}}
&{{2t_it_j+t_i^2t_j^2-2t_j^2-2t_i^2+1}\over
{(1-t_j^2)(1-t_i^2)(1-t_it_j)^3}} \cr
\endpmatrix
\eqno(4.21)
$$
for $i\not=j$. The leading powers of the elements of $\Delta_m^{(1)}$,
as $t_1\to\infty$, are
$$
\Delta_m^{(2)}=\pmatrix
1/t_1^2&0&1/t_1&1/t_1&\dots&1/t_1&1/t_1 \cr
0&1/t_1^6&1/t_1^3&1/t_1^3&\dots&1/t_1^3&1/t_1^3 \cr
1/t_1&1/t_1^3&*&*&\dots&*&* \cr
1/t_1&1/t_1^3&*&*&\dots&*&* \cr
\dots&\dots&\dots&\dots&\dots&\dots&\dots \cr
1/t_1&1/t_1^3&*&*&\dots&*&* \cr
1/t_1&1/t_1^3&*&*&\dots&*&* \cr
\endpmatrix
\eqno(4.22)
$$
where *'s stand for the terms of the order of $O(1)$.
Therefore  
$$
\det\Delta_m=O\left({1\over {t_1}^8}\right)\,,\qquad t_1\to\infty.
\eqno(4.23)
$$
By (4.17),
$$
\det\De_m\sim {\const.C(t_1,\dots, t_m)\over t_1^8},\qquad
t_1\to\infty, 
$$
hence $C(t_1,\dots,t_m)$ is constant in $t_1$. The same argument on
$t_2,\dots,t_m$  
shows that it is independent of $t_2,\dots,t_m$, so it is indeed a
constant, say $C_m$, i.e., 
$$
\det\Delta_m={C_m{\prod_{1 \le i<j\le m}(t_i-t_j)^8}\over
{\prod_{i=1}^m(1-t_i^2)^4\prod_{1 \le i<j \le m}(1-t_it_j)^8}} 
\eqno(4.24)
$$
To prove that $C_m=1$, let us consider the
asymptotic behavior of $\det\Delta_m$ as $t_1\to 1$ with
$t_2,\dots,t_m$ fixed and close to zero. Then on the one hand,
we have from (4.28) that
$$
\lim_{t_1 \to 1}(1-t_1^2)^4\det\Delta_m={C_m\over
C_{m-1}}\det\Delta_{m-1}, 
\eqno(4.25)
$$
On the other hand,
$$
\Delta_m^{(1)}=\pmatrix
{1\over (1-t_1^2)}&0&*&\dots&0 \cr
0&{1\over (1-t_1^2)^3}&*&\dots&* \cr
*&*& & & \cr
\dots&\dots& &\Delta_{m-1}&  \cr
*&* & & &  \cr
\endpmatrix
\eqno(4.26)
$$
where the terms $*$ are regular at $t_1=1$. Hence the leading term of
the Laurent series of $\det \De_m$ at $t_1=1$ is $(1-t_1)^{-4}\det
\De_{m-1}$, which shows that
$$
{C_m\over C_{m-1}}\det\Delta_{m-1}=\det\Delta_{m-1}
\eqno(4.27)
$$
Thus $C_m=C_{m-1}$. Repeating this argument we get that 
$$
C_m=C_{m-1}=\dots=C_2=C_1=1
$$
Therefore $C_m=1$. Proposition 4.1 is proved.

Similarly we prove the following proposition.

{\bf Proposion 4.3.} {\it 
$$
\Omega_m=\pmatrix
\omega_{11}&\omega_{12}&\dots&\omega_{1m} \cr
\omega_{21}&\omega_{22}&\dots&\omega_{2m} \cr
\dots&\dots&\dots&\dots \cr
\omega_{m1}&\omega_{m2}&\dots&\omega_{mm} \cr
\endpmatrix
\eqno(4.28)
$$
with
$$
\omega_{ii}={(1-t_i^2)^3\prod_{j\not =i}(1-t_it_j)^4\over 
\prod_{j\not=i}(t_i-t_j)^4}
\eqno(4.29)
$$
and
$$
\omega_{ij}={(1-t_i^2)^2(1-t_j^2)^2
\prod_{k\not=i}(1-t_it_k)^2\prod_{k\not=j}(1-t_jt_k)^2\over(1-t_it_j)
\prod_{k\not=i}(t_i-t_k)^2\prod_{k\not=j}(t_j-t_k)^2},\qquad
{i \not= j}.   
\eqno(4.30)
$$}

Put now
$$
t_i=\tanh(\pi s_i) \qquad i=1,\dots,m.
\eqno(4.31)
$$
Then by (4.29) and (4.30),
$$
\omega_{ij}=(1-t_i^2)^{3/2}(1-t_j^2)^{3/2}
\frac{\prod_{k\not=i}\coth^2\pi(s_i-s_k)\prod_{k\not=j}
\coth^2\pi(s_j-s_k)}
{\cosh\pi(s_i-s_j)}.
\eqno(4.32)
$$
In addition, by (4.7),
$$
\det\Delta_m=
\frac{\prod_{1\le i<j \le m}\tanh^8\pi(s_i-s_j)}
{\prod_{i=1}^m(1-t_i^2)^4}\,.
\eqno(4.33)
$$
By (2.14),
$$
k_m(s_1,\dots,s_m)={K_m(t_1,\dots,t_m)\over\prod_{i=1}^m p(t_i)}\,,
\quad p(t)=\frac{1}{\pi(1-t^2)}\,.
\eqno(4.34)
$$
Now let us substitute $y_i$'s in (4.6) by
$$
\frac{y_i}{(1-t_i^2)^{3/2}}\,, \qquad i=1,\dots,m;
\eqno(4.35)
$$
then
$$\eqalign{
K_m(t_1,\dots,t_m)
&={\prod_{i=1}^mp(t_i)\over2^m\prod_{1 \le  i<j \le m}
\tanh^4\pi(s_i-s_j)}\cr
&\times\int_{-\infty}^{\infty}\dots\int_{-\infty}^{\infty}
|y_1\dots y_m|e^{-{1\over 2}(Y\Sigma_m,Y)}\,dy_1\dots dy_m \cr}
\eqno(4.36)
$$
where
$$\Sigma_m=\pmatrix
\sigma_{11}&\sigma_{12}&\dots&\sigma_{1m} \cr
\sigma_{21}&\sigma_{22}&\dots&\sigma_{2m} \cr
\dots&\dots&\dots&\dots \cr
\sigma_{m1}&\sigma_{m2}&\dots&\sigma_{mm} \cr
\endpmatrix
\eqno(4.37)
$$
with
$$
\sigma_{ij}={\prod_{k\not=i}\coth^2\pi(s_i-s_k)
\prod_{k\not=j}\coth^2\pi(s_j-s_k)\over
\cosh\pi(s_i-s_j)}\,. 
\eqno(4.38)
$$
Now substitute $y_i$'s in the formula (4.36) by
$$
\frac{y_i}{\prod_{j\not= i}\coth^2\pi(s_i-s_j)}\,,\qquad i=1,\dots,m;
\eqno(4.39)
$$
then
$$
K_m(t_1,\dots,t_m)
=2^{-m}\prod_{i=1}^mp(t_i)
\int_{-\infty}^{\infty}\dots\int_{-\infty}^{\infty}
|y_1\dots y_m|e^{-{1\over 2}(Y\Gamma_m,Y)}\,dy_1\dots dy_m,
\eqno(4.40)
$$
where
$$
\Gamma_m=\left({1\over \cosh\pi (s_i-s_j)}\right)_{i,j=1}^m
\eqno(4.41)
$$
Thus
$$
k_m(s_1,\dots,s_m)
=2^{-m}\int_{-\infty}^{\infty}\dots\int_{-\infty}^{\infty}
|y_1\dots y_m|e^{-{1\over 2}(Y\Gamma_m,Y)}\,dy_1\dots dy_m.
\eqno(4.42)
$$
Theorem 1.2 is proved.

{\it Proof of Theorem 1.3.} Let
$$
f_n(t)=c_0+c_1t+\dots +c_{n-1}t^{n-1}+c_nt^n,\qquad
g_n(t)=c_0t^n+c_1t^{n-1}+\dots + c_{n-1}t+c_n.
\eqno (4.43)
$$
Then if $\t_k\not= 0$ is a zero of $f_n(t)$ then $\t_k^{-1}$ is a zero
of $g_n(t)$. Hence if $1<a<b$ then
$$
\xi_f(a,b)=\xi_g(a^{-1},b^{-1}),
\eqno (4.44)
$$
where $\xi_f(a,b)$ is the number of zeros of $f_n(t)$ in the interval
$(a,b)$. Observe that the distribution of the vector of coefficients
$(c_0,\dots,c_n)$ 
coincides with the one of the vector $(c_n,\dots, c_0)$. Hence
$$\eqalign{
\E [\xi_f(a_1,b_1)\dots \xi_f(a_m,b_m)]&=
\E [\xi_g(a_1^{-1},b_1^{-1})\dots \xi_g(a_m^{-1},b_m^{-1})]\cr &= 
\E [\xi_f(a_1^{-1},b_1^{-1})\dots \xi_f(a_m^{-1},b_m^{-1})].\cr}
\eqno (4.45)
$$
Take $a_j=t_j$ and $b_j=t_j+\De t_j$, $j=1,\dots,m,$ and get $\De
t_j\to 0$. Since 
$$
|a^{-1}-b^{-1}|=a^{-2}|a-b|+O(|a-b|^2),\qquad a\to b,
$$
 we deduce that
$$
{K_{nm}(t_1^{-1},\dots, t_m^{-1})\over t_1^2\dots t_m^2}=
K_{nm}(t_1,\dots,t_m).
\eqno (4.46)
$$
Hence
$$
{K_{nm}(t_1^{-1},\dots, t_m^{-1})\over p(t_1^{-1})\dots p(t_m^{-1})}=
{K_{nm}(t_1,\dots, t_m)\over p(t_1)\dots p(t_m)},
\eqno (4.47)
$$
because
$$
p(t^{-1})={1\over \pi|1-t^{-2}|}=t^2p(t).
$$
This proves that
$$
k_{nm}^{\out}(s_1,\dots,s_m)=k_{nm}(s_1,\dots,s_m).
$$
Theorem 1.3 is proved.

{\it Proof of Theorem 1.4.} Let $|t_1|,\dots,|t_l|<1$ and
$|t_{l+1}|,\dots, |t_m|>1$. Denote by
$D_{nm}(x_1,y_1,\dots,x_m,y_m;t_1,\dots t_m)$ the joint distribution
density of the vector 
$$
(f_n(t_1),f_n(t_1),\dots,f_n(t_l),f'_n(t_l),
g_n(t_{l+1}),g'_n(t_{l+1}),\dots g_n(t_m),g_n'(t_m)).
$$
Then
$$\eqalign{
K_{nm}(t_1,\dots,t_m)&=\int_{-\infty}^\infty\dots \int_{-\infty}^\infty
|y_1\dots y_m| \cr
&\times D_{nm}(0,y_1,\dots, 0,y_m;t_1,\dots,t_l,t_{l+1}^{-1},
\dots, t_m^{-1})dy_1\dots dy_m.\cr}
\eqno (4.48)
$$
The covariance
$$
\E f_n(t_i)g_n(t_j^{-1})=t_i^n+t_i^{n-1}t_j^{-1}+\dots +t_i
t_j^{1-n}+t_j^{-n},\qquad 1\le i\le l<j\le m, 
$$
goes to 0 as $n\to\infty$, together with the partial derivatives in
$t_i,t_j$, while
$$\eqalign{
&\lim_{n\to\infty} \E f_n(t_i)f_n(t_j)={1\over 1-t_it_j}\,,\qquad 1\le
i,j\le l,\cr
&\lim_{n\to\infty} \E g_n(t_i^{-1})g_n(t_j^{-1})
={1\over 1-t_i^{-1}t_j^{-1}}\,,\qquad l+1\le
i,j\le m.\cr}
\eqno (4.49)
$$
This proves that the limiting Gaussian kernel
$D_m=\lim_{n\to\infty} D_{nm}$ is factored,
$$\eqalign{
D_m(x_1,y_1,\dots x_m,y_m;t_1,\dots, t_l,&t_{l+1}^{-1},\dots, t_m^{-1})
= D_l(f)(x_1,y_1,\dots,x_l,y_l;t_1,\dots,t_l)\cr
&\times D_{m-l}(g)(x_{l+1},y_{l+1},\dots, x_m,y_m;t_{l+1}^{-1},\dots,
t_m^{-1}).\cr}
$$
Hence the correlation function $K_m$ is also factored. Theorem 1.4 is
proved.

\beginsection 5. Correlations between Zeros of the SO(2) 
Random Polynomial \par

In the following two sections we discuss the correlation
functions of real zeros and the variance of the number of real zeros
of the SO(2) random polynomial.   Let $f_n(t)$ be a SO(2) random
polynomial, that 
is 
$$
f_n(t)=\sum_{k=0}^n c_kt^k, \eqno (5.1)
$$
where $\{ c_k\}$  are real independent Gaussian
random variables with 
$$
\E c_k=0,\qquad \E c_k^2=\sg_k^2=
\pmatrix
n\\ k
\endpmatrix\,.
\eqno (5.2)
$$
In this case (1.4) reduces to
$$\eqalign{
A_n(t)&=(1+t^2)^n,\cr
B_n(t)&=nt(1+t^2)^{n-1},\cr  
C_n(t)&=n(1+nt^2)\,(1+t^2)^{n-2},\cr}
\eqno (5.3)
$$
which gives
$$
p_n(t)={\sqrt n\over \pi(1+t^2)}\,
\eqno (5.4)
$$
(see [BS], [EK], [BBL1], and others). 
The average value of real zeros is
$$
\E\,\#\{\,k\:\t_k\in\R\,\}=\int_{-\infty}^\infty p_n(t)\,dt=\sqrt n\,,
\eqno (5.5)
$$
and the normalized density,
$$
{p_n(t)\over \sqrt n}={1\over \pi(1+t^2)}\,,
\eqno (5.6)
$$
is the Cauchy distribution density. Observe that both (5.4)
and (5.5) are exact relations for all $n$. Let us compute the
two-point correlation function $K_{n2}(t_1,t_2)$. 

Define
$$
\f_n(t)={f_n(t)\over (1+t^2)^{n/2}}\,.
\eqno (5.7)
$$
Then the real zeros of $\f_n(t)$ coincide with those of
$f_n(t)$. Hence, similar to (2.6), we can write that
$$
K_{n2}(t_1,t_2)=\int_{-\infty}^\infty\int_{-\infty}^\infty
|y_1y_2|D_{n2}(0,y_1,0,y_2;t_1,t_2)dy_1dy_2,
\eqno (5.8)
$$
where $D_{n2}(x_1,y_1,x_2,y_2;t_1,t_2)$ is the distribution density
function of the vector 
$$
(\f_n(t_1),\, \f'_n(t_1),\,
\f_n(t_2),\,\f'_n(t_2)).
$$ 
Observe that
$$
\E \f_n(t_1)\f_n(t_2)=\rho_n(t_1,t_2)\equiv
{(1+t_1t_2)^n\over (1+t_1^2)^{n/2}(1+t_2^2)^{n/2}},
\eqno (5.9)
$$
and
$$\eqalign{
& \E \f'_n(t_1)\f_n(t_2)={\partial\rho_n\over \partial t_1}(t_1,t_2)=
n\rho_n\,{(t_2-t_1)\over (1+t_1t_2)(1+t_1^2)}\,;\cr
& \E \f'_n(t_1)\f'_n(t_2)={\partial^2\rho_n\over \partial t_1
\partial t_2 }(t_1,t_2)\cr
&=
-n^2\rho_n\,{(t_2-t_1)^2\over (1+t_1t_2)^2(1+t_1^2)(1+t_2^2)}
+n\rho_n\,{1\over (1+t_1t_2)^2}\,.\cr}
\eqno (5.10)
$$
Define the random variable $\psi_n(t)$ as
$$
\psi_n(t)={\f_n'(t) (1+t^2)\over \sqrt n}\,.   
\eqno(5.11)
$$
Let $D_{n2}(x_1,y_1,x_2,y_2;t_1,t_2)$ be the distribution density
function of the random vector
$(\f_n(t_1),\psi_n(t_1),\f_n(t_2),\psi_n(t_2))$. Then after a
change of variables, (5.8) is rewritten as
$$
{K_{n2}(t_1,t_2)\over p_n(t_1)p_n(t_2)}
=\pi^2\int_{-\infty}^{\infty}\int_{-\infty}^{\infty}|y_1y_2|
D_{n2}(0,y_1,0,y_2;t_1,t_2)\,dy_1dy_2 .
\eqno(5.12)
$$
From (5.10),
$$\eqalign{
& \E \psi_n(t_1)\f_n(t_2)=
\rho_n\,{\sqrt n(t_2-t_1)\over (1+t_1t_2)}\,;\cr
& \E \psi_n(t_1)\psi_n(t_2)=
\rho_n\left[-{n(t_2-t_1)^2\over (1+t_1t_2)^2}
+{(1+t_1^2)(1+t_2^2)\over (1+t_1t_2)^2}\right]\,,\cr}
\eqno (5.13)
$$
hence the covariance matrix of the vector
$(\f_n(t_1),\psi_n(t_1),\f_n(t_2),\psi_n(t_2))$ is
$$
\Delta_n=
\pmatrix
1 & 0 & \rho_n  & \rho_n\,{\sqrt n(t_1-t_2)\over (1+t_1t_2)}\\
0 & 1 &  \rho_n\,{\sqrt n(t_2-t_1)\over (1+t_1t_2)} &
\rho_n a \\
\rho_n &  \rho_n\,{\sqrt n(t_2-t_1)\over (1+t_1t_2)}
& 1 & 0 \\
\rho_n\,{\sqrt n(t_1-t_2)\over (1+t_1t_2)} & 
\rho_n a & 0 & 1\\
\endpmatrix
\eqno(5.14)
$$
where
$$
a=a(t_1,t_2)=-{n(t_2-t_1)^2\over (1+t_1t_2)^2}
+{(1+t_1^2)(1+t_2^2)\over (1+t_1t_2)^2}\,.
\eqno (5.15)
$$
Suppose that $t_1$ and $t_2$ are two distinct fixed points. Then as
$n\to\infty$, the quantity
$$
|\rho_n|=\left[ 1-{(t_1-t_2)^2\over
(1+t_1^2)(1+t_2^2)}\right]^{n/2}
\eqno (5.16)
$$
goes to 0 exponentially fast, and hence $\De_n$ approaches the unit
matrix exponentially fast. By (5.12) this implies that
$$
\lim_{n\to\infty} {K_{n2}(t_1,t_2)\over p_n(t_1)p_n(t_2)}=1,
$$
and the rate of convergence is exponential. In the same way we obtain
that if $t_1,\dots,t_m$ are $m$ distinct fixed points then 
$$
{K_{nm}(t_1,\dots,t_m)\over p_n(t_1)\dots p_n(t_m)}
=\pi^m\int_{-\infty}^{\infty}\dots \int_{-\infty}^{\infty}|y_1\dots
y_m| 
D_{nm}(0,y_1,\dots 0,y_m;t_1,\dots, t_m)\,dy_1\dots dy_m ,
\eqno(5.17)
$$
where $D_{nm}(x_1,y_1,\dots x_m,y_m;t_1,\dots, t_m)$ is a 
$(2m)\times (2m)$ Gaussian
density with the  covariance matrix
$$
\De_n=(\De^{(n)}_{ij})_{i,j=1,\dots, m},
$$
where
$$
\De^{(n)}_{ii}=
\pmatrix
1 & 0 \\
0 & 1 
\endpmatrix,\qquad
\De^{(n)}_{ij}=\rho_n(t_i,t_j)
\pmatrix
 1  & 
{\sqrt n(t_i-t_j)\over (1+t_it_j)}\\
{\sqrt n(t_j-t_i)\over (1+t_1t_2)} &
 a(t_i,t_j)
\endpmatrix
\eqno (5.18)
$$
where $\rho_n(t_1,t_2)$ and $a(t_1,t_2)$ are defined in (5.9) and
(5.15), respectively. For fixed
different $t_1,\dots, t_m$ the matrix $\De_n$ approaches the unit
matrix exponentially fast, and this implies that
$$
\lim_{n\to\infty} {K_{nm}(t_1,\dots,t_m)\over p_n(t_1)\dots
p_n(t_m)}=1,
$$
and the rate of convergence is exponential. This means the
independence of the distribution of real zeros at distinct fixed
points. 

The formula (5.17) is simplified if we make the change of variables
$t=\tan \th$ (stereographic projection). Consider therefore the random
function 
$$
g_n(\th)=\sum_{j=0}^n c_k\sin^k(\th)\cos^{n-k}(\th), \qquad -{\pi\over
2}<\th\le {\pi\over 2}\,. 
\eqno (5.18)
$$
Then
$$
g_n(\th)=\cos^n\th \, f_n(\tan\th),
\eqno (5.19)
$$
hence if $\{\t_j\}$ are zeros of $f_n(t)$ then 
$$
\{\eta_j=\arctan \t_j\}
\eqno (5.20)
$$
are  zeros of $g_n(\th)$. When $c_n=0$, $g_n(\th)$ has an extra zero
$\eta=\pi/2$. Since the probability that $c_n=0$ is equal to zero, the
joint probability distributions of zeros of the functions $g_n(\th)$
and $f_n(\tan\th)$ coincide. By (5.6) the zeros $\{\eta_j\}$ are
uniformly distributed on the interval $[-\pi/2,\pi/2]$. The function
$g_n(\th)$ is a Gaussian 
random function with zero mean and the covariance function
$$
\E g_n(\th_1)g_n(\th_2)=\sum_{k=0}^n\pmatrix
n\\ k\endpmatrix \sin^k\th_1\cos^{n-k}\th_1\sin^k\th_2\cos^{n-k}\th_2
=\cos^n(\th_1-\th_2).
\eqno (5.21)
$$
The function $g_n(\th)$ is periodic of period $\pi$ if $n$ is even, and
it is periodic
of period $2\pi$ if $n$ is odd. It is convenient to consider
$g_n(\th)$ on the circle $S^1=\R/(2\pi)\Z$ of the length $2\pi$. This
circle is the covering space of the original circle
$\R/\pi\Z$. If $g_n(\th)=0$ then $g_n(\th+\pi)=0$ as well. On $S^1$  
the distribution of $g_n(\th)$ is invariant with respect to the
shift 
$$
\th\to\a+\th\mod 2\pi,
$$
hence it is SO(2)-invariant.

Let $K_{nm}(\th_1,\dots,\th_m)$ be the correlation function of the
zeros $\{\eta_j\}$ of $g_n(\th)$. Assume that 
$$
\th_j-\th_k\not=0\mod\pi,\qquad 1\le j,k\le m.
\eqno (5.22)
$$
Then (5.17) gives that
$$
{K_{nm}(\th_1,\dots,\th_m)\over (\pi^{-1}\sqrt n)^m}
=\pi^m\int_{-\infty}^{\infty}\dots \int_{-\infty}^{\infty}|y_1\dots
y_m| 
D_{nm}(0,y_1,\dots 0,y_m;\th_1,\dots, \th_m)\,dy_1\dots dy_m ,
\eqno(5.23)
$$
where $D_{nm}(x_1,y_1,\dots x_m,y_m;\th_1,\dots, \th_m)$ is a 
$(2m)\times (2m)$ Gaussian
density with the  covariance matrix
$$
\De_n=(\De^{(n)}_{ij})_{i,j=1,\dots, m},
$$
where
$$
\De^{(n)}_{ij}=\cos^n(\th_i-\th_j)
\pmatrix
 1  & 
\sqrt n \tan(\th_i-\th_j)\\
\sqrt n \tan(\th_j-\th_i) &
-n\tan^2(\th_i-\th_j)+\cos^{-2}(\th_i-\th_j)
\endpmatrix
\eqno (5.24)
$$
Observe that $D_{nm}(x_1,y_1,\dots x_m,y_m;\th_1,\dots, \th_m)$ is the
probability distribution density of the vector
$$
\left(g_n(\th_1), {g_n'(\th_1)\over\sqrt n},\dots,g_n(\th_m),
{g_n'(\th_m)\over \sqrt n}\right), 
$$
and it is nondegenerate provided that $n\ge 2m-1$ and (5.22) holds.

Consider now the scaling limit of the correlation functions. The
straightened zeros are
$$
\z_j={\eta_j \sqrt n\over \pi}.
$$
They are uniformly distributed on the circle of the length $2\sqrt n$. 
The limit $m$-point correlation function of $\{\z_j\}$ is
$$
k_{m}(s_1,\dots,s_m)=\lim_{n\to\infty}
{K_{nm}(\th_1,\dots,\th_m)\over (\pi^{-1}\sqrt n)^m}\,,
\qquad
\th_i={ s_i\pi \over \sqrt n},
\eqno (5.25)
$$
Let us find $k_m(s_1,\dots,s_m)$. We have that
$$
\lim_{n\to\infty}\cos^n\left( {(s_i-s_j)\pi\over \sqrt n}\right)
=e^{-\pi^2(s_i-s_j)^2/2},
\eqno (5.26)
$$
and by (5.24),
$$
\lim_{n\to\infty}\De_n=
\De=(\De_{ij})_{i,j=1,\dots,m}
$$ 
with
$$
\De_{ij}=e^{-\pi^2(s_i-s_j)^2/2}
\pmatrix
1  & \pi(s_i-s_j) \\
\pi(s_j-s_i) & 1-\pi^2(s_i-s_j)^2
\endpmatrix
\eqno (5.27)
$$ 
By (5.23) this gives that
$$
k_m(s_1,\dots,s_m)
=\pi^m\int_{-\infty}^{\infty}\dots \int_{-\infty}^{\infty}
|y_1\dots y_m| 
d_{m}(0,y_1,\dots 0,y_m;s_1,\dots, s_m)\,dy_1\dots dy_m ,
\eqno(5.28)
$$
and $d_{m}(x_1,y_1,\dots x_m,y_m;s_1,\dots, s_m)$ is a Gaussian
density with the covariance matrix $\De$.
Let $\Om$ be $m\times m$ matrix obtained by deleting all odd rows and
odd columns from the matrix $\De^{-1}$. Then we can write (5.28) as
$$
k_m(s_1,\dots,s_m)
={1\over 2^m(\det \De)^{1/2}}
\int_{-\infty}^{\infty}\dots \int_{-\infty}^{\infty}|y_1\dots
y_m|e^{-{1\over 2}(Y\Omega,Y)}
\,dy_1\dots dy_m.
\eqno(5.29)
$$
In Appendix C below we prove that
$$
\De>0 \quad \iff\quad s_i\not= s_j,\; 1\le i<j\le m,
\eqno (5.30)
$$
so the formula (5.29) is well-defined when the points $s_i$ are
distinct.

For $m=2$, (5.29) reduces to
$$
k_2(s_1,s_2)
={1\over 4\pi^2(\det \De)^{1/2}}
\int_{-\infty}^{\infty}\int_{-\infty}^{\infty}|y_1
y_2|e^{-{1\over 2}(Y\Omega,Y)}
\,dy_1dy_2.
\eqno(5.31)
$$
where
$$
\De=
\pmatrix 
1 & 0 & e^{-s^2/2} &  se^{-s^2/2} \\
0 & 1 & -se^{-s^2/2} & (1-s^2)e^{-s^2/2}
 \\
e^{-s^2/2} & -se^{-s^2/2} & 1 & 0 \\
se^{-s^2/2} & (1-s^2)e^{-s^2/2}
 & 0 & 1 
\endpmatrix\,,\qquad s=\pi(s_1-s_2),
\eqno (5.32)
$$
and 
$$
\Om=(\De^{-1})_{\{2,4\}},
$$
i.e., $\Om$ is the $2\times 2$-submatrix of $\De^{-1}$ at the second
and the fourth rows and columns. Observe that
$$
\det\Om={\det \De_{\{1,3\}}\over\det\De}={1-e^{-\pi^2s^2}\over \det\De}\,.
$$
A direct computation gives 
$$
\det\De=(1-e^{-s^2})^2-s^4e^{-s^2}
\eqno (5.33)
$$
and
$$
\Om= \pmatrix
A&B\\
B&A\\
\endpmatrix
$$
where
$$
A={1-e^{-s^2}-s^2e^{-s^2}\over \det\De}     ,\qquad
B={e^{-s^2/2}(e^{-s^2}+s^2-1)\over \det\De}.
\eqno (5.34)
$$
Since 
$$
\int_{-\infty}^{\infty}\int_{-\infty}^{\infty}
|y_1y_2|e^{{-{1\over 2}}(Ay_1^2+Ay_2^2+2By_1y_2)}\,dy_1dy_2
={4\over\det\Om}\left(1+{\de\over\sqrt{1-\de^2}}
\arcsin\de\right),
$$
where $\de=B/A$ (see Appendix A), we obtain from (5.25) that
$$\eqalign{
k_2(s_1,s_2)&={1\over (\det\De)^{1/2}\det\Om}
\left(1+{\de\over\sqrt{1-\de^2}}
\arcsin\de\right)\cr
&={(\det\De)^{1/2}\over 1-e^{-s^2}}
\left(1+{\de\over\sqrt{1-\de^2}}
\arcsin\de\right)\cr
&={[(1-e^{-s^2})^2-s^4e^{-s^2}]^{1/2}\over 1-e^{-s^2}}
\left(1+{\de\over\sqrt{1-\de^2}}
\arcsin\de\right),
\cr}
\eqno(5.35)
$$
where $s=\pi(s_1-s_2)$ and
$$
\de={e^{-s^2/2}(e^{-s^2}+s^2-1)
\over 1-e^{-s^2}-s^2e^{-s^2}}
\eqno (5.36)
$$
It is worth to note that Hannay [Han] has calculated the limit
two-point correlation function of zeros for the complex random SU(2)
polynomial, 
and our calculation of (5.35) is very similar to the one of
Hannay. 

As $s\to 0$,
$$
\det \De={s^8\over 12}+O(s^{10}),\qquad \de=1-{s^2\over 6}+O(s^4),
$$
which implies that
$$
k_2(s_1,s_2)={\pi^2|s_1-s_2|\over 4}+O(|s_1-s_2|^2),\qquad s_1-s_2\to
0. 
\eqno (5.37)
$$
As $s\to\infty$,
$$
\det \De=1-s^4e^{-s^2}+O(e^{-s^2}),\qquad
\de=-s^2e^{-s^2/2}+O(e^{-s^2/2}), 
$$
which implies that
$$
k_2(s_1,s_2)=1+{\pi^4(s_1-s_2)^4e^{-\pi^2(s_1-s_2)^2}\over
2}+O((s_1-s_2)^2e^{-\pi^2(s_1-s_2)^2}),\qquad |s_1-s_2|\to\infty. 
\eqno (5.38)
$$
Thus we have proved the following theorem.

{\bf Theorem 5.1.} {\it Let $\{\tau_j\}$ be zeros of a
random SO(2) polynomial $f_n(t)$ of degree $n$, and
$$
\z_j={\sqrt n\, \arctan \tau_j \over \pi }.
$$
be the straightened zeros.  Then the limit $m$-point
correlation function $k_m(s_1,\dots,s_m)$ of $\{\z_j\}$ is given by
the formula (5.29) where $\De$ is a $(2m)\times(2m)$ symmetric matrix
which consists  of $2\times 2$ blocks $\De_{ij}$ defined in
(5.26), and $\Om=(\De^{-1})_{\text{\rm even}}$ is the $m\times m$
matrix of the elements of $\De^{-1}$ with even indices. The 2-point
correlation function is given by the formula 
(5.35), and its asymptotics as $s_1-s_2\to 0$ and
$|s_1-s_2|\to\infty$ are given in (5.37) and (5.38), respectively.}

The graph of $k_2(0,s)$ is shown in Fig.2.

\centinsert{\pscaption{\psboxto (6in;2.4in){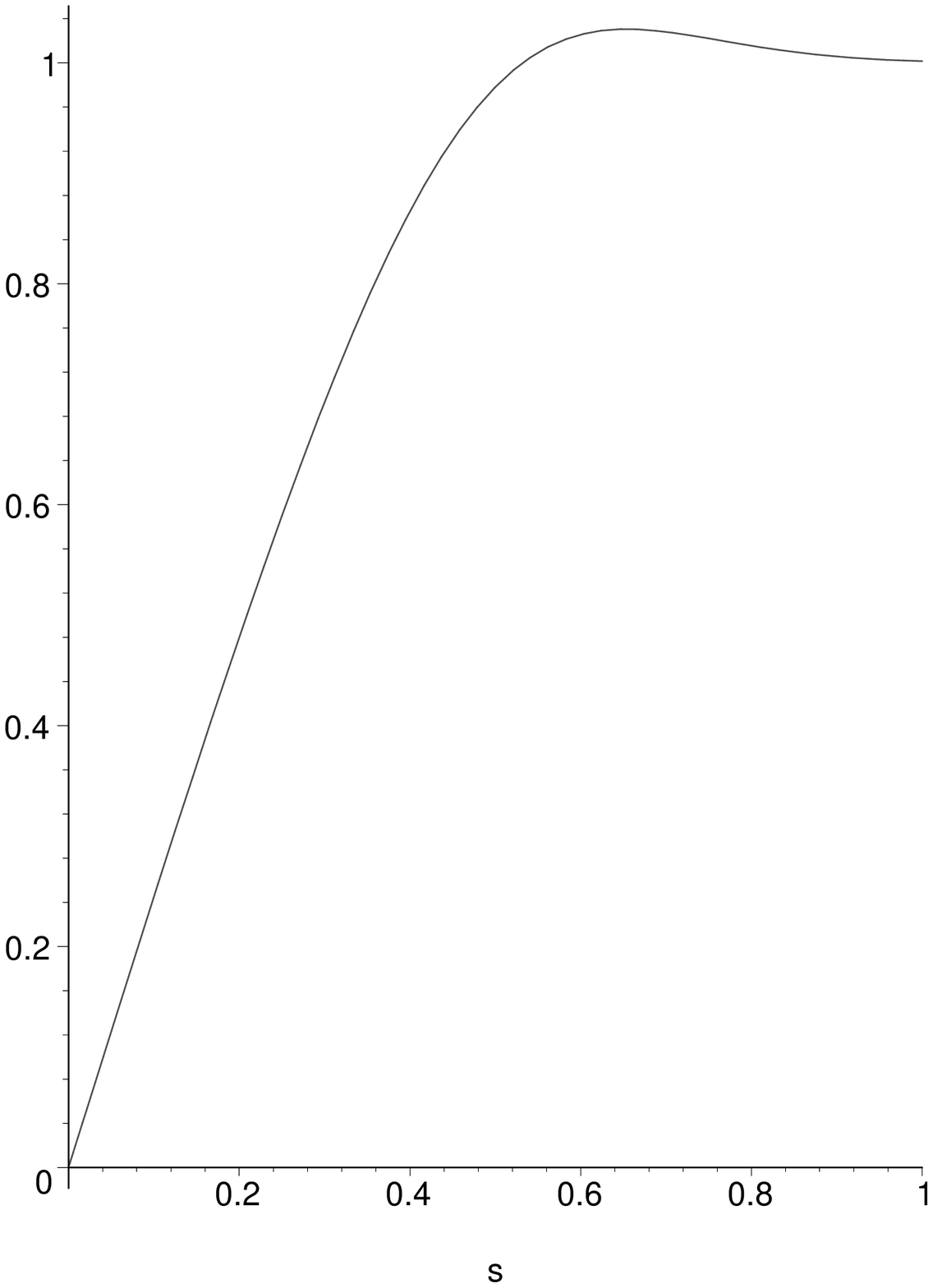}}
{\srm Fig 2:  The two-point correlation function of straightened zeros of
the SO(2) random polynomial. }}

\beginsection 6. Variance of the number of real zeros of the SO(2)
random polynomial \par 

Here we calculate the variance of the random variable $\xi_n(a,b)$ in
the case when $f_n(t)$ is the SO(2) random polynomial. By definition,
$$
\Var\xi_n(a,b)=\E\xi_n^2(a,b)-(\E \xi_n(a,b))^2.
\eqno(6.1)
$$
By (2.17),  
$$
\E\xi_n^2(a,b)=\int_a^bp_n(t)dt+\int_a^b\int_a^bK_{n2}(t_1,t_2)dt_1dt_2.
\eqno(6.2)
$$
Since
$$
(\E \xi_n(a,b))^2=\int_a^b p_n(t_1)dt_1\int_a^bp_n(t_2)dt_2,
\eqno(6.3)
$$
we obtain that
$$\eqalign{
\Var\xi_n(a,b)
&=\int_a^bp_n(t)dt+
\int_a^b\int_a^b[K_{n2}(t_1,t_2)-p_n(t_1)p_n(t_2)]dt_1dt_2\cr
&=\int_a^bp_n(t)dt+
\int_a^b\int_a^b\left[{K_{n2}(t_1,t_2)\over p_n(t_1)p_n(t_2)}-1\right]
p_n(t_1)p_n(t_2)dt_1dt_2\cr}
\eqno(6.4)
$$
When $t_1$ and $t_2$ are separated, the difference
$$
{K_{n2}(t_1,t_2)\over p_n(t_1)p_n(t_2)}-1
$$ 
is exponentially small,
hence the main contribution to the last integral comes from close
$t_1,t_2$. Let us put 
$$
t_1=t,\qquad t_2=t+{s\over p_n(t)}.
$$
Then
$$
\int_a^b\int_a^b\left[{K_{n2}(t_1,t_2)\over p_n(t_1)p_n(t_2)}-1\right]
p_n(t_1)p_n(t_2)dt_1dt_2\sim \int_a^b p_n(t) dt\int_{-\infty}^\infty
(k_2(0,s)-1)ds,
$$
hence
$$
\Var \xi_n(a,b)\sim \int_a^b p_n(t) dt\left[ 1-\int_{-\infty}^\infty
(1-k_2(0,s))ds\right].
$$
Thus
$$
\Var \xi_n(a,b)\sim C\sqrt n\int_a^b {dt\over \pi(1+t^2)},
$$
where
$$
C=1-\int_{-\infty}^\infty
(k_2(0,s)-1)ds,
$$
and $k_2(s_1,s_2)$ is the two-point correlation fumction given in
(5.35). In particular,
$$
\Var \xi_n(-\infty,\infty)\sim C\sqrt n.
$$
Numerical value of $C$ is
$$
C=0.5717310486902\dots.
$$

\beginsection Appendix A. Calculation of an Integral \par

In this appendix we show that
$$
\int_{-\infty}^\infty\int_{-\infty}^\infty|y_1y_2|e^{-{1\over
2}(Ay_1^2+Cy_2^2+2By_1y_2)}dy_1dy_2={4\over AC-B^2}\left(1+{\de\over
\sqrt{1-\de^2}} \arcsin\de\right),
\eqno(A.1)
$$
where $\de=B/\sqrt{AC}$.
By a change of variables we can reduce the integral (A.1) to 
$$
I=\int_{-\infty}^\infty\int_{-\infty}^\infty|y_1y_2|e^{-{1\over 2}(y_1^2+y_2^2+2\delta y_1y_2)}dy_1dy_2
$$
Changing then
$$
\eqalign{
y_1&={1\over\sqrt{2}}(x_1+x_2), \cr
y_2&={1\over\sqrt{2}}(x_1-x_2), \cr}
$$
we obtain that
$$
I={1\over 2}\int_{-\infty}^\infty\int_{-\infty}^\infty|x_1^2-x_2^2|
e^{-{1\over2}[(1+\delta)x_1^2+(1-\delta)x_2^2]}dx_1dx_2.
$$
Let 
$$\eqalign{
x_1&={r\over \sqrt{1+\delta}}\cos\theta, \cr
x_2&={r\over \sqrt{1-\delta}}\sin\theta. \cr}
$$
Then
$$
\eqalign{
I&={1\over 2(1-\delta^2)^{1/2}}\int_0^\infty r^3e^{-{r^2\over 2}}dr
\int_0^{2\pi}\left|{{\cos^2\theta}\over{1+\delta}}
-{{\sin^2\theta}\over{1-\delta}}\right|
d\theta \cr
&={1\over{(1-\delta^2)^{3/2}}}\int_0^{2\pi}|(1-\delta)
\cos^2\theta-(1+\delta)\sin^2\theta|d\theta \cr
&={1\over{(1-\delta^2)^{3/2}}}\int_0^{2\pi}|
\de-\cos 2\theta|d\theta \cr}
$$
Evaluating the last integral we obtain that
$$
I={4\over(1-\delta^2)^{3/2}}(\sqrt{1-\delta^2}+\delta\arcsin\delta)
$$
hence (A.1) is proved.

\beginsection Appendix B. Positivity of the Covariance Matrix 
\par

Assume that $f_n(t)=c_0+c_1t+\dots+c_nt^n$ is a polynomial with
independent random coefficients such that $\E c_k=0$ and $0<\E
c^2_k<\infty$. We show in this appendix that the covariance matrix of
the random vector 
$$
F_n=(f_n(t_1),f'_n(t_1),\dots, f_n(t_m),f'_n(t_m))
\eqno (\text{ B}.1)
$$
is positive, if $t_1,\dots,t_m$ are pairwise different
and $n\ge 2m-1$. Consider the
real-valued quadratic form associated with the covariance matrix,
$$\eqalign{
Q(\a,\b)&=\sum_{j,k} \bigl[
\a_j\a_k \E f_n(t_j) f_n(t_k)
+\a_j\b_k \E f_n(t_j) f_n'(t_k)\cr
&+\b_j\a_k \E f'_n(t_j)f_n(t_k)+\b_j\b_k
\E f'_n(t_j) f_n(t_k)\bigr]\cr
&=\E\left|\sum_j\bigl [ \a_j f_n(t_j)+\b_j f'_n(t_j)\bigr]\right|^2\cr
&=\sum_{k=0}^n\sg_k^2
\left[\sum_{j=1}^m\left( \a_jt^k_j+\b_jkt_j^{k-1}\right)\right]^2\cr
}$$
The generalized Vandermonde matrix 
$$
\bigl (t^k_j,kt^{k-1}_j\bigr )_{j=1,\dots, m;\; k=0,\dots, 2m-1}
$$
is nondegenerate, hence
$$
\sum_{k=0}^n\sg_k^2
\left[\sum_{j=1}^m\left( \a_jt^k_j+\b_jkt_j^{k-1}\right)\right]^2>0,
$$
provided that not all $\a_j,\b_j$ are zero. This proves that the
quadratic form $Q(\a,\b)$ is positive, hence the covariance matrix of
the random vector (B.1) is positive as well. 

The proof remains valid if $c_k$ are, in general, dependent random
variables with 
positive covariance matrix $(V_{kl}=\E c_kc_l)_{k,l=0,\dots,n}$. In
this case
$$
Q(\a,\b)=\sum_{k,l=0}^n V_{kl} d_kd_l>0,
\qquad d_k=\sum_{j=1}^m\left(
\a_jt^k_j+\b_jkt_j^{k-1}\right),
$$
provided that $n\ge 2m-1$ and not all  $\a_j,\b_j$ are zero. 

\beginsection Appendix C. Proof of the Inequality $\De>0$ \par

Let $\De=(\De_{jk})_{j,k=1,\dots, m}$ where
$$
\De_{jk}=e^{-(s_j-s_k)^2/2}
\pmatrix
1  & (s_j-s_k) \\
(s_k-s_j) & 1-(s_j-s_k)^2
\endpmatrix
$$
We prove in this appendix that $\De>0$. Consider the complex-valued
quadratic form 
associated with $\De$,
$$
Q(\a,\b)=\sum_{j,k} e^{-(s_j-s_k)^2/2}\bigl[ \a_j\ov{\a}_k
+(s_j-s_k)\a_j\ov{\b}_k +(s_k-s_j)\b_j\ov{\a}_k
+[1-(s_j-s_k)^2]\b_j\ov{\b}_k\bigr].
$$
Using the formulae
$$\eqalign{
e^{-(s_j-s_k)^2/2}&={1\over \sqrt{2\pi}}\int_{-\infty}^\infty
e^{i(s_j-s_k)x}e^{-x^2/2}dx,\cr
(s_j-s_k)e^{-(s_j-s_k)^2/2}&={1\over \sqrt{2\pi}}\int_{-\infty}^\infty
e^{i(s_j-s_k)x}(-ix)e^{-x^2/2}dx,\cr
[1-(s_j-s_k)^2]e^{-(s_j-s_k)^2/2}&={1\over \sqrt{2\pi}}\int_{-\infty}^\infty
e^{i(s_j-s_k)x}x^2e^{-x^2/2}dx,\cr}
$$
we can rewrite $Q$ as
$$\eqalign{
Q(\a,\b)&={1\over \sqrt{2\pi}}\int_{-\infty}^\infty dx e^{-x^2/2}
\sum_{j,k} \bigl[ \a_j\ov{\a}_ke^{i(s_j-s_k)x}
+\a_j\ov{\b}_ke^{i(s_j-s_k)x}(-ix)\cr
& +\b_j\ov{\a}_ke^{i(s_j-s_k)x}(ix)
+\b_j\ov{\b}_ke^{i(s_j-s_k)x}x^2\bigr].\cr}
$$
Define
$$
f(x)=\sum_{j=1}^m \a_j e^{is_j x},\qquad
g(x)=\sum_{j=1}^m \b_j e^{is_j x}.
$$
Then 
$$
Q(\a,\b)={1\over \sqrt{2\pi}}\int_{-\infty}^\infty 
|f(x)+ixg(x)|^2 e^{-x^2/2}dx.
$$
The function $f(x)+ix g(x)$ is not identically zero
provided that not all $\a_j,\b_j$ are 0. Indeed, for every test
function $\f(x)$,
$$
\int_{-\infty}^\infty (f(x)+ixg(x))\f(x)dx
=\sum_{j=1}^m (\a_j\tilde\f(s_j)+\b_j\tilde\f'(s_j)),
$$
where $\tilde\f(s)$ is the Fourier transform of $\f(x)$. We can
localize the function $\tilde\f(s)$ near $s_j$ and make the last sum
nonzero. This proves that $f(x)+ix g(x)$ is not identically zero, and
hence $Q(\a,\b)>0$.
Hence $\De>0$. 

\vskip 1cm

{\it Acknowledgement.} This work was supported in part by the National
Science Foundation, Grant No. DMS--9623214, and this support is
gratefully acknowledged. 

\vskip 1cm 

\beginsection  References \par

\item{[BS]} A.T. Bharucha-Reid and M. Sambandham, {\it Random
polynomials,} {  Academic Press,} { New York,} 1986. 

\item {[BP]} A. Bloch and G. Polya, {\it On the roots of certain
algebraic equations,} Proc. London Math. Soc. (3) {\bf 33} (1932),
102-114.

\item {[BBL1]} E. Bogomolny, O. Bohigas and P. Leboeuf, {\it
Distribution 
of roots of random polynomials,} Phys. Rev. Lett. {\bf 68} (1992),
2726-2729. 

\item{[BBL2]} E. Bogomolny, O. Bohigas and P. Leboeuf, {\it Quantum
chaotic dynamics and random polynomials.} Preprint, 1996.
 
\item {[Dys]} F. J. Dyson,
{\it Correlation between eigenvalues of a random matrix,}
Commun. Math. Phys. {\bf 19} (1970), 235--250. 

\item{[EK]} A. Edelman and E. Kostlan, {\it  How many zeros of a random
polynomial are real?} { Bull. Amer. Math. Soc.,} {\bf 32} (1995)
1-37. 

\item {[EO]} P. Erd\"os  and A. Offord, {\it On the number of real
roots of random algebraic equation,} Proc. London Math. Soc. {\bf 6}
(1956), 139--160. 

\item {[ET]} P. Erd\"os and P. Tur\'an,
{\it On the distribution of roots of polynomials,} Ann. Math. {\bf 51}
(1950), 105--119.

\item {[Han]} J. H. Hannay, {\it Chaotic analytic zero points: exact
statistics for those of a random spin state,} J. Phys. A:
Math. Gen. {\bf 29} (1996), 101-105.

\item {[IM]} I. A. Ibragimov and N. B. Maslova,
{\it On the average number of real zeros of random polynomials. I,
II,} 
Teor. Veroyatn. Primen. {\bf 16} (1971), 229--248, 495--503. 

\item{[K1]} M. Kac, {\it On the average number of real roots of a
random 
algebraic equation,} { Bull. Amer. Math. Soc.} {\bf 49} (1943),
314-320.

\item{[K2]} M. Kac, {\it On the average number of real roots of a
random 
algebraic equation. II,} { Proc. London Math. Soc.} {\bf
50} (1948), 390-408. 

\item{[K3]} M. Kac, {\it Probability and related topics in physical
sciences.} 
Wiley (Interscience), { New York}, 1959. 

\item {[Kos]} E. Kostlan,
{\it On the distribution of roots of random polynomials}, From
Topology to 
Computation: Proceedings of the Smalefest (M. W. Hirsh, J. E. Marsden, 
and M. Shub, eds.), Springer-Verlag, New York, 1993, pp. 419--431.

\item {[Leb1]} P. Leboeuf, 
{\it Phase space approach to quantum dynamics}, 
J. Phys. A: Math. Gen. {\bf 24} (1991), 4575.

\item {[Leb2]} P. Leboeuf, {\it Statistical theory of chaotic
wavefunctions: a model in terms of random analytic functions,}
Preprint, 1996.
 
\item {[LS]} P. Leboeuf and P. Shukla, {\it Universal Fluctuations of
zeros of Chaotic Wavefunctions,} Preprint, 1996.
 
\item {[LO]} J. Littlewood and A. Offord, {\it On the number of real
roots of random algebraic equation. I,II,III,} J. London
Math. Soc. {\bf 13} (1938), 288-295; Proc. Cambr. Phil. Soc. {\bf 35}
(1939), 133--148; Math. Sborn. {\bf 12} (1943), 277--286.

\item {[LS]} B. F. Logan and L. A. Shepp,
{\it Real zeros of a random polynomial. I, II, } Proc. London
Math. Soc. {\bf 18} (1968), 29--35, 308--314.

\item {[Maj]} E. Majorana,
Nuovo Cimento {\bf 9} (1932), 43--50. 

\item {[Mas1]} N. B. Maslova, {\it On the variance of the number of
real roots of random polynomials,} Teor. Veroyatn.
Primen. {\bf 19} (1974), 36--51.
 
\item {[Mas2]} N. B. Maslova, {\it On the distribution of the number
of real roots of random polynomials,} Teor. Veroyatn.
Primen. {\bf 19} (1974), 488--500.
 
\item {[Meh]} M. L. Mehta, {\it Random matrices}, Academic Press,New
York, 1991. 

\item{[M-A]} G. Mezincescu, D. Bessis, J. Fournier, G. Mantica, and
F. D. Aaron, {\it  Distribution of roots of random real polynomials.}
Preprint, 1996. 

\item {[SS]} M. Shub and S. Smale,
{\it Complexity of Bezout's Theorem II: Volumes and Probabilities},
Computational Algebraic Geometry (F. Eysette and A. Calligo, eds.),
Progr. Math., vol. 109, Birkhauser, Boston, 1993, pp. 267--285. 

\item {[Ste]} D. Stevens,
{\it Expected number of real zeros of polynomials,} Notices
Amer. Math. Soc. {\bf 13} (1966), 79. 




\end